\pdfoutput=1
\documentclass[preprint]{elsarticle}

\usepackage[utf8]{inputenc}

\usepackage{amsmath,amssymb}

\usepackage{hyperref}
\usepackage{booktabs}

\journal{International Journal of Electrical Power \& Energy Systems}
\renewcommand{\Re}{\operatorname{Re}}
\renewcommand{\Im}{\operatorname{Im}}

\newcommand{\Npv}{\ensuremath{n_{\text{PV}}}}
\newcommand{\Npq}{\ensuremath{n_{\text{PQ}}}}

\newcommand{\sumij}{\ensuremath{\sum_{\substack{ij\\j<i}}}}
\DeclareMathOperator{\sgn}{sgn}

\begin{document}

\begin{frontmatter}

  \title{A Padé-Weierstrass technique for the rigorous enforcement
    of control limits in power flow studies}

  % Possible alternative titles:
  % (1) A Padé-Weierstrass technique for solving the problem of enforcing control
  %     limits in power flow studies
  % (2) A Padé-Weierstrass technique for solving the problem of
  %     limits in power flow studies
  % (3) A Padé-Weierstrass technique for the rigorous enforcement
  %     of control limits in power flow studies
  % (4) Rigorous solution to the problem of limits in power flow using a
  %     new Padé-Weierstrass technique

  \author{A.~Trias\corref{corr1}}
  \ead{triast@aia.es}
  \cortext[corr1]{Corresponding author}
  
  \author{J. L.~Mar\'in\corref{corr2}}
  \ead{marinjl@aia.es}
  \cortext[corr2]{Principal corresponding author}

  \address{Aplicaciones en Informática Avanzada S.L.\fnref{website},
    Edificio ESADECREAPOLIS, Av.~de la Torre Blanca 57, 08172 Sant Cugat
    del Vallès, Barcelona, Spain}

  \fntext[website]{Company website: \url{http://www.aia.es}}

  \begin{abstract}
    A new technique is presented for solving the problem of enforcing control
    limits in power flow studies. As an added benefit, it greatly increases the
    achievable precision at nose points. The method is exemplified for the case
    of Mvar limits in generators regulating voltage on both local and remote
    buses. Based on the framework of the Holomorphic Embedding Loadflow Method
    (HELM), it provides a rigorous solution to this fundamental problem by
    framing it in terms of \emph{optimization}. A novel Lagrangian formulation
    of power-flow, which is exact for lossless networks, leads to a natural
    physics-based minimization criterion that yields the correct solution. For
    networks with small losses, as is the case in transmission, the AC power
    flow problem cannot be framed exactly in terms of optimization, but the
    criterion still retains its ability to select the correct solution. This
    foundation then provides a way to design a HELM scheme to solve for the
    minimizing solution. Although the use of barrier functions evokes interior
    point optimization, this method, like HELM, is based on the analytic
    continuation of a germ (of a particular branch) of the algebraic curve
    representing the solutions of the system. In this case, since the constraint
    equations given by limits result in an unavoidable singularity at $s=1$,
    direct analytic continuation by means of standard Padé approximation is
    fraught with numerical instabilities.  This has been overcome by means of a
    new analytic continuation procedure, denominated Padé-Weierstrass, that
    exploits the covariant nature of the power flow equations under certain
    changes of variables. One colateral benefit of this procedure is that it can
    also be used when limits are not being enforced, in order to increase
    the achievable numerical precision in highly stressed cases.
  \end{abstract}

  \begin{keyword}
    % IEEEkeywords:
    Load flow analysis \sep power system analysis computing \sep power system
    modeling \sep power system simulation \sep power engineering computing
    %% PACS codes: (https://publishing.aip.org/publishing/pacs/pacs-2010-regular-edition)
    % 88.80.hh 	Transmission grids
    % 88.80.hm 	AC power transmission
    % 84.30.Bv 	Circuit theory
    % 02.60.Cb 	Numerical simulation; solution of equations
    % 02.60.Gf 	Algorithms for functional approximation
    \PACS 88.80.hh \sep 88.80.hm \sep 84.30.Bv \sep 02.60.Cb \sep 02.60.Gf
    
    %% MSC codes: (http://www.ams.org/mathscinet/msc/msc2010.html)
    \MSC[2010] 14H50 \sep 14H81 \sep 30B10 \sep 30B40 \sep 30B70 \sep 30E10
    \sep 94C99
  \end{keyword}

\end{frontmatter}

\section{Introduction}
\label{sec:intro}

The so-called power flow (or load flow) problem consists in finding the steady
state of a power network where sources and loads are specified in terms of
constant power. It is one of the cornerstones of power systems analysis, upon
which many other analyses and tools are built. Mathematically, the core problem
consists in a system of nonlinear algebraic equations for the bus voltages,
given simply by the power balance equations at every bus. There are however
additional controls, mostly dealing with voltage regulation, that need to be
taken into account. The most important and ubiquitous among these is voltage
regulation by synchronous generators, exerted either locally or at a remote
bus. Other important examples are under-load tap changers (ULTC) in
transformers, automatically switched shunt capacitor/reactor banks, HVDC line
controls, and modern FACTS devices. In all these cases the control device can be
represented as one or more additional \emph{equality constraint} equations
representing how the regulated magnitude is to be restricted; and
correspondingly, new variables (which can be thought of as the regulating
resources) are added to the system in order to keep a balance with between the
total numbers of equations and unknowns. The problem is that all controls have
resource limits: generators have reactive Mvar limits, ULTC have a limited
range, etc.  When a regulating resource hits a limit, the behavior of the
control changes: the regulated magnitude is now allowed to vary from the
commanded setpoint, while the resource remains at its saturation value. This
switching behavior turns the power flow into a hard computational
problem, particularly for large and complex network models.

An early review by Stott~\cite{Stott74} on powerflow methods discusses the two
main approaches for enforcing limits, which are still used in most iterative
power flow engines today (for a more recent review,
see~\cite{GomezExpAlvarado09}). One straightforward method consists in
``type-switching'', where the equations change as needed: the controlling
parameter (e.g. the Q injection for a generator; or the tap ratio for an ULTC)
is initially an independent variable; but if, after solving the power flow, it
hits a limit, it becomes a fixed parameter and the corresponding constraint
equation is removed. This method requires solving different power flows as the
switching takes place.  The other strategy is based on using ``feedback
adjustments''~\cite{AllanArruda82,ChangBrandwajn88}: the controlling resource is
considered always a parameter, but it is modified on-the-fly between iterations,
according to how the controlled voltage deviates from the setpoint. This has the
advantage that the equations remain fixed, so that the iterative scheme is
simpler.

However, both approaches face the problem of slow convergence due to
``rebounds'': when a given control saturates during an intermediate step of
these methods, it may later require to be reverted back to the unsaturated
state.  This happens because of nonlinear interactions among controls,
particularly when several of them are close to their limits. Sometimes these
rebounds enter into cycles, preventing convergence~\cite{GomezExpAlvarado09}.
Many methods have been developed to avoid this sort of
interactions~\cite{ChangBrandwajn91,Vlachogiannis94}, but their justification
remains heuristic. To this day, power flow practitioners still have to rely on a
combination of personal experience and heuristics in order to deal with these
convergence problems~\cite{PowerWorld04,Siemens12}, which usually constitute a
time-consuming task.  This state of affairs is somewhat surprising since
transmission planners always need to deal with peak cases, which are quite close
to feasibility boundaries and contain a large percentage of generators (more
than 30\% is not unusual) saturated at their Mvar limits.

There have been many studies characterizing the effects of limits in the
transition to voltage
collapse~\cite{DobsonLu1992,VuLiu92,HiskensChakrabarti96,YueVenkatasubramanian07,KataokaShinoda05,ZhaoZhang08},
but none of these have resulted in a practical power flow method.  In order to
find further insights with actual application to the general power flow problem,
one needs to look into the related areas of Optimal Power Flow (OPF). Many
authors have approached the problem of calculating maximum loadability in the
presence of limits using OPF techniques~\cite{Irisarri97}, but it was
specifically the representation of control limits by means of
\emph{complementarity constraints}~\cite{Vournas00,RomanRosehart05,Rosehart05}
that triggered new approaches to the general power flow~\cite{Kataoka05,Zhao08},
most notably~\cite{Murray2015}.  One key idea in these papers is the realization
that a power flow with control limits can be tackled as a constrained
optimization problem, even though its aim is not computing a loadability
limit. Of course this begs the question of what \emph{objective function} to
use, and in our opinion none of these works justify their choices
satisfactorily.  Having said that, using at least \emph{some} minimization
criterion seems a more principled approach than just using traditional
adjustment heuristics or type-switching, since neither one of those techniques
allow for a characterization of the solution they arrive to. One should be aware
that there could be in general more than one valid power flow solution,
differing in how control devices end up either in their saturated or
un-saturated state.

This paper presents a general-purpose power flow method in which the problem of
enforcing control limits is solved rigorously, based on a minimization principle
that stems from the fundamental physics of electrical networks. The method is
developed for the case of Mvar limits in generators regulating voltage on both
local and remote buses. Since it is based on the framework of the Holomorphic
Embedding Loadflow Method (HELM)~\cite{Trias12,Trias15,TriasMarin16}, the method
also inherits the properties of being constructive and therefore produces
unequivocal results: it yields the correct operating solution when the case is
feasible, and it detects infeasibility otherwise.

The paper is structured as follows. Section~\ref{sec:theproblem} presents the
problem definition and lays down the nomenclature and conventions used in the
method. Section~\ref{sec:Lagrangian} develops a Lagrangian formulation of power
flow problem, starting from the nonlinear DC problem (not to be confused with
the linear ``DC'' approximation to AC problems) and extending it to lossless AC
networks. This provides the minimization criterion underlying the proposed
method, since it shows how to understand the complementarity constraints in the
framework of a constrained minimization procedure. Section~\ref{sec:HELMscheme}
then describes how the HELM scheme is implemented for this problem. In contrast
to barrier methods used in traditional nonlinear programming (such as Interior
Point Methods), the HELM technique is purely based on power series and their
analytic continuation. This section shows how to accommodate the constraint
equations and, more importantly, how to ensure that the embedding guarantees
that the HELM reference germ is consistent with reactive limits and voltage
setpoints. Having laid down the essentials of the method, Section~\ref{sec:PW}
then deals with a technical obstacle in the analytic continuation: the
unavoidable singularity found at $s=1$, which provokes a very slow convergence
of the Padé approximants. This non-trivial numerical problem has been overcome
thanks to a new technique based on certain changes of variables having the
property that they leave the embedded power flow equations formally invariant
under the change. This allows to construct the analytic continuation, still
based on Padé approximants, incrementally in stages, similarly to the classic
Weierstrass idea of analytic continuation along a path. Since this technique is
absolutely crucial to the success of practical implementations on a computer,
the method has been named after it. Finally, Section~\ref{sec:NumResults} shows
a sample of numerical results.  Appendix Section~\ref{app:methsummary}
recapitulates all steps of the method in summarized form.

% Key ideas for the intro:
% \begin{itemize}
% \item Observing control limits in power flow studies is a problem that is still
%   not solved rigorously. Many heuristic approaches but no rigorous theoretical
%   treatment.
% \item Defining the problem. The issue of insufficient information: concurrent
%   controls, conflicting controls, competing saturations, priorities, etc.
%   Dynamics info is absent.  One should still try to make the best use of
%   available information. How? This paper is one proposal.
% \item Quick overview of the state of the art approches.  How limits are
%   enforced in traditional LF methods: type-switching vs.\ on-the-fly
%   adjustments.
% \item Unconventional approaches coming from the field of OPF, particularly
%   from studies of the maximum loadability limit (a.k.a. transfer
%   limits). Relaxations of the LF problem into Semidefinite Programming
%   problems. The encoding of limits (which are inequality constraints) as
%   ``complementarity conditions''.
% \end{itemize}

\section{The problem}
\label{sec:theproblem}

Let us establish the nomenclature by considering a general power system
comprised of constant-power injections $S_i=P_i+jQ_i$ and constant-current
injections $I_i$. The power flow equations are given by the current balance at
each bus,
\begin{equation}
  \label{eq:PF}
  \sum_{j=0}^n y_{ij} (V_i-V_j) + Y_i^\text{sh} V_i = I_i + \frac{S_i^*}{V_i^*}
  \quad (i=1, \ldots, n),
\end{equation}
where $V_i$ are bus voltages, $y_{ij}$ are branch admittances, and
$Y_i^\text{sh}$ are shunt admittances to ground, which are assumed to group
contributions from line charging susceptances, shunts from transformer modeling,
shunt reactor/capacitor banks, and any loads modeled as a constant
admittance. As it is commonplace in power systems, the \emph{active} sign
convention will be used here, so that generators inject positive current $I_i$
and power $P_i$, while loads inject negative values for those same variables.
The network is composed of a total of $n+1$ buses where, without loss of
generality, a swing bus is assumed at index 0 (and a swing voltage $V_0$ should
be specified). It is also customary to define the transmission admittance matrix
as:
\begin{equation*}
  Y_{ij} \equiv
  \begin{cases}
    \sum_{k=0}^n y_{ik} &\text{if } i=j \\
    -y_{ij} & \mbox{if } i \ne j
  \end{cases}
  \quad , 
\end{equation*}
so that the powerflow equations can be expressed in a more compact form:
\begin{equation}
  \label{eq:PF_allPQ}
  \sum_{j=0}^n Y_{ij} V_j + Y_i^\text{sh} V_i = I_i + \frac{S_i^*}{V_i^*}
  \quad (i=1, \ldots, n).
\end{equation}

These are the power flow equations in their most fundamental form, where all bus
injections on the r.h.s.\ are specified. Since all magnitudes are
complex-valued, system~\eqref{eq:PF_allPQ} consists of $2n$ equations in $2n$
unknowns, the complex voltages $V_i$. In order to contemplate voltage regulation
or any other type of control, one includes additional \emph{constraint}
equations and variables to~\eqref{eq:PF_allPQ}, always obeying the basic
algebraic rule that the number of independent equations and the number of
variables has to match---otherwise one ends up with an over/under-determined
system.

\subsection{Voltage regulation (PV buses) with no Q-limits}
\label{subsec:PVunlim}
In the following, the formulation will focus on the most ubiquitous type of
control: the automatic voltage regulation (AVR) performed by generators.  This
kind of control may be either local or remote. When it is local to the bus, the
bus is commonly referred to as ``PV-type''.  Whether local or remote, for the
purposes of steady-state power flow, the control materializes as an additional
equation constraining the voltage modulus of the controlled bus $i$,
$|V_i|=V_i^\text{sp}$ (the setpoint). Correspondingly, the reactive power
injection $Q_j$ of the controlling generator $j$ becomes a new variable in the
system. If there are several generators regulating the same voltage, some
sharing factors need to be used in order to obtain a single effective degree of
freedom, to avoid under-determinacy.

Therefore, when the controlling resources (in this case, generator Mvars) are
assumed unlimited, the system of power flow equations simply gets augmented with
new equations (equality constraints) and new variables. In the holomorphic
embedding method, unlimited controls are easily incorporated into the
formulation, after taking proper care of embedding the new constraints in a way
that preserves both holomorphicity and consistency with the reference state at
$s=0$. This can be done in a different number of ways, as shown
in~\cite{Subramanian2013,Trias15,Wallace2016}.  For the treatment of standard PV
buses, our favored embedding scheme is as follows:
\begin{align}
  \label{eq:PF_PQ}
  i \in \text{PQ:} & \quad \sum_{j=0}^n Y_{ij} V_j(s) + sY_i^\text{sh} V_i(s) =
    sI_i + \frac{sS_i^*}{V_i^*(s^*)} \\
  \label{eq:PF_PV}
  k \in \text{PV:} &  \quad \sum_{j=0}^n Y_{kj} V_j(s) + sY_k^\text{sh} V_k(s) =
    sI_k + \frac{sP_k-jQ_k(s)}{V_k^*(s^*)} \\
  \label{eq:PVulimconstraint}
  k \in \text{PV:} &  \quad V_k(s)V_k^*(s^*) =
                     1 + s\left( W_k^\text{sp} - 1\right) .
\end{align}
Here $W_k^\text{sp}$ is defined as $\left(V_k^\text{sp}\right)^2$. The swing
voltage is assumed to be $V_0=1$; this can be achieved either by performing a
global normalization of variables, or more simply by embedding the swing as
$V_0(s)=1+s(V_0-1)$.

Considering the real and imaginary parts of voltages, this is a system of
$2\Npq+3\Npv$ equations, in $2\Npq+3\Npv$ unknowns. However, these equations can
be simplified and the dimensionality reduced to $2\Npq+\Npv$ equations and
unknowns, as shown in \ref{app:methsummary}.

\subsection{Dealing with Q-limits: complementarity equations}
\label{subsec:PVlim}

The focus is now switched to the problem of resource limits, which is the main
subject of the paper. When a generator reaches either limit, $Q^\text{min}$ or
$Q^\text{max}$, its controlling injection saturates at that limit, and the
controlled voltage can no longer be sustained at the setpoint. Moreover the sign
of the control error is important, as it should be consistent with the sign of
the control sensitivity: if the generation saturates at $Q^\text{max}$, the
achieved voltage should be $V\le V^\text{sp}$, while if it saturates at
$Q^\text{min}$ one should have $V^\text{sp}\le V$. One should be careful with
the special case where the PV bus condition has reversed sensibility, which
corresponds to a power flow scenario on the ``wrong side'' of the generator's
Q-V curve. Although this is a mathematically valid power flow solution, it
cannot be considered operationally correct, as it makes voltage regulation
unstable. Further, it can be shown how, from the point of view of either HELM or
a continuation power flow, these anomalous cases graze a point of collapse
somewhere along the path. In the following, we will assume that the user has
corrected these anomalies beforehand, so that all generator voltage controls
have the correct sensitivity, $\partial V/\partial Q > 0$.

Given the complementary nature of the constraints, they may be expressed through
the equations:
\begin{align}
  \label{eq:complementarity}
  &\Bigl( V_iV_i^* - W^\text{sp} \Bigr) \Bigl( Q^\text{max}_i - Q_i \Bigr)
  \Bigl( Q_i - Q^\text{min}_i \Bigr) = 0 \\
  \label{eq:Qlims}
  &\Bigl( Q^\text{max}_i - Q_i\Bigr) \ge 0; \qquad
   \Bigl( Q_i - Q^\text{min}_i \Bigr) \ge 0
\end{align}
plus the additional consistency requirements regarding voltage control
sensitivities:
\begin{equation}
  \label{eq:sensitiv_cond}
  \begin{split}
  Q_i=Q^\text{max}_i &\implies |V_i| \le V^\text{sp} \\ 
  Q_i=Q^\text{min}_i &\implies |V_i| \ge V^\text{sp}  
  \end{split}
\end{equation}

The problem now is finding an equation or set or equations which, when
appropriately embedded, would yield a solution satisfying all of these
conditions~\eqref{eq:complementarity}--\eqref{eq:sensitiv_cond} at $s=1$. In
contrast with the unlimited case, where the equality
constraint~\eqref{eq:PVulimconstraint} can be embedded in quite a
straightforward way, the limited case contains a number of deep subtleties that
make it much more non-obvious.  Attempting an ad-hoc embedding
of~\eqref{eq:complementarity} with no regard to the other conditions is likely
to end up in a faulty HELM scheme.  In the following section, it will be
demonstrated how, thanks to a suitable Lagrangian formulation of the power flow
problem, this problem can be correctly framed as one of \emph{minimization}, and
thus obtain a HELM scheme that properly enforces all of the above equality and
inequality constraints.

%%%% EMBEDDING OF THE SWING VOLTAGE
% Recall that HELM's reference solution is strategically chosen to be the
% zero-injection state, where all buses are energized at the swing voltage
% value $V_\text{sw}$. For convenience, all voltages can be normalized to this
% voltage (or, equivalently, the swing can be embedded as
% $1+s(V_\text{sw}-1)$), so that at zero order one has $V_i[0]=1$ for for all
% buses and $Q_i[0]=0$ for all PV buses.

%%% CHARACTERIZATION OF THE SOLUTION
% What is the nature of the solution that equation~\eqref{eq:compl_embed},
% together with the choice of the HELM reference state, generate?  Given the
% fact that, in general, there could be more than one feasible selection of
% saturated and non-saturated generators, we would need to characterize this
% particular solution obtained by the method.

\section{Lagrangian formulation of power flow}
\label{sec:Lagrangian}

This section shows how the power flow equations can be derived from a Lagrangian
formulation, much in the spirit of classical mechanics. This is then used to
show how the problem of limits can be tackled in terms of minimization, and from
there devise a suitable HELM scheme.  Other authors have also approached the
same problem using OPF-like minimization techniques~\cite{Murray2015}, but here
the treatment specifically addresses the holomorphic embedding formulation. In
particular, it will be shown to have an intriguing resemblance to interior point
methods, but with the advantage of remaining a method based on holomorphic
functions and their analytic continuation.

\subsection{(Nonlinear) DC power flow}

In order to show the essential points of the Lagrangian formulation, the
attention is first turned to the pure DC problem, where equations are simpler.
The nonlinear power flow equations of an all-DC network may be written as:
\begin{equation}
  \label{eq:DC_PF}
  \sum_{j=0}^n g_{ij} (V_i-V_j) + G_i^\text{sh} V_i = I_i + \frac{P_i^*}{V_i^*}
  \quad (i=1, \ldots, n),
\end{equation}
This is \emph{not} the commonly used ``DC power flow'' approximation to AC;
rather, it is the exact DC counterpart to the AC problem~\eqref{eq:PF}, where
voltages, conductances, and power and current injections are all real
magnitudes. Systems of this type are not common in utility grids, but they
do appear for instance in spacecraft, shipboard systems, and some modern
microgrids~\cite{TriasMarin16}. In any case, this system will be shown to be
quite useful as it clarifies up many issues that are otherwise obscured in the
AC formulation.

It is straightforward to verify that equations~\eqref{eq:DC_PF} can be derived
from the minimization of the following Lagrangian:
\begin{equation}
  \label{eq:DCLagrangian}
  \mathcal{L} = \frac{1}{2} \sumij g_{ij} \left( V_i - V_j \right)^2
  +\frac{1}{2} \sum_i G_i^\text{sh} V_i^2 - I_i V_i - P_i \ln V_i  \;,
\end{equation}
where the double summation of the first term avoids counting links twice, but
does include the swing (at index $j=0$); while the single summation of the rest
of the terms runs only over non-swing buses. For now the term Lagrangian is used
here to refer to what is essentially a potential energy function (in the spirit
of classical physics), but concepts of constrained optimization and
Lagrangian duality will be used shortly after.

The fact that the powerflow equations admit a description in terms of a
potential function is remarkable and should not be dismissed as a mere
curiosity.  It follows that the powerflow problem can be seen as equivalent to a
problem of finding static equilibria in classical mechanics. It is thus
instructive to analyze the physical meaning of each term in this Lagrangian, as
it can provide fruitful insights. The first term is the sum of the power losses
($I^2R$) over all transmission branches of the network (divided by two).  The
next term accounts for the power consumption due to shunt conductances. The next
term, $I_iV_i$, is the power supplied by constant-current injections at each
bus. The last term, $P_i \ln V_i$, is also power supplied into the network,
although the particular form of its expression will become clearer shortly.
Reading~\eqref{eq:DCLagrangian} as a problem in classical mechanics, one could
envision a hypothetical system made up of masses connected with springs and
sitting under some peculiar ``gravitational'' fields. Conductances $g_{ij}$ and
$G_i^\text{sh}$ would play the role of spring constants, while the rest of the
terms could be interpreted as some sort of on-site potential producing a force
field acting locally on each bus. It is also interesting to note that the
operational solution corresponds to a stable equilibrium of the system (a
minimum), while all other power flow solutions (``black
branches''~\cite{Trias12}) correspond to unstable equilibria (saddle points),
although this will not be proven here.

The role of constant power injections becomes clearer
when~\eqref{eq:DCLagrangian} is viewed as resulting from a slightly different,
\emph{constrained} minimization problem:
\begin{equation}
  \label{eq:DCminimiz}
  \begin{split}
    \mathcal{E} \left( \{V_i\} \right) \equiv \frac{1}{2} &\sumij g_{ij}
    \left(V_i-V_j\right)^2 + \frac{1}{2}\sum_i G_i^\text{sh} V_i^2 - I_i V_i \\
    \underset{\{V_i\}}{\text{minimize:}} &\quad \mathcal{E}(\{V_i\}) \\
    \text{subject to:} &\quad V_k - V_k^\text{sp} = 0
    \qquad \left( k\in \{\text{V-regulated}\} \right)
  \end{split}
\end{equation}
Here the voltage is regulated to a setpoint $V_k^\text{sp}$ on those buses for
which there was a non-zero constant-power injection specified in
problem~\eqref{eq:DC_PF}.  Since negative voltages are not allowed, these
constraint equations can be written equivalently as
$\ln V_k^\text{sp}-\ln V_k=0$. This constrained problem is then solved by the
standard method of Lagrange multipliers, using the following Lagrangian:
\begin{equation}
  \label{eq:DCLagrangian2}
  \mathcal{L} \equiv \mathcal{E}
  \; - \sum_k P_k \ln \left( \frac{V_k}{V_k^\text{sp}} \right)  \;,
\end{equation}
which is the same as~\eqref{eq:DCLagrangian}, except for additional constant
terms $P_i\ln V_i^\text{sp}$ which do not depend on the voltages. The
stationarity conditions then take the same form as the original power flow
equations~\eqref{eq:DC_PF}. Therefore, constant power injections can be seen as
the Lagrange multipliers of minimization
problem~\eqref{eq:DCminimiz}. Additionally, this view has the advantage of
having a more direct physical interpretation in terms of a fundamental energy
balance, since now the primal minimization problem is on $\mathcal{E}$: power
losses, plus power consumption, minus power supply.  Under this light, the
standard power flow problem~\eqref{eq:DC_PF}, in which constant power injections
(both load and generation) are specified, could be interpreted as minimizing the
transmission network losses. This is a key insight and a powerful guide when
approaching the AC case.

Adding now resource constraints to the regulation resources:
\begin{equation}
  \label{eq:DCminLimited}
  \begin{split}
    \underset{\{V_i\}}{\text{minimize:}} &\quad \mathcal{E}\left( \{V_i\} \right) \\
    \text{subject to:} &\quad V_k - V_k^\text{sp}= 0
    \qquad \left( k\in \{\text{V-regulated}\} \right) \\
    &\quad P_k^\text{min} \le P_k \le P_k^\text{max}
  \end{split}
\end{equation}
Note how these inequality constraints are on the dual variables of
problem~\eqref{eq:DCminimiz}. Let us solve the maximization of the dual
problem by means of logarithmic barrier methods, by defining:
\begin{equation*}
  \mathcal{B}_\mu \left( \{P_k\} \right) \equiv \mathcal{E}
   - \! \sum_k P_k \ln \left( \frac{V_k}{V_k^\text{sp}} \right)
  + \mu \ln \left(P_k^\text{max}-P_k\right)
  + \mu \ln \left(P_k-P_k^\text{min}\right)  
\end{equation*}
where $\mu>0$ is the barrier parameter that is made to converge to zero. The
stationarity conditions for maximizing $\mathcal{B}_\mu$ yield:
\begin{equation}
  \label{eq:DCBarrier}
  - \ln \left( \frac{V_k}{V_k^\text{sp}} \right)
  - \frac{\mu}{P_k^\text{max}-P_k} + \frac{\mu}{P_k-P_k^\text{min}} = 0
\end{equation}
Rearranging,
\begin{equation*}
  \left(\ln V_k-\ln V_k^\text{sp}\right)
  \left(P_k^\text{max}-P_k\right) \left(P_k-P_k^\text{min}\right) =
  \mu \left( P_k^\text{max} + P_k^\text{min} - 2P_k \right)
\end{equation*}
In the limit $\mu\rightarrow0$, this equation is essentially the same as the
complementarity constraint~\eqref{eq:complementarity} (when translated to the DC
problem), since $\ln V_k=\ln V_k^\text{sp}$ if and only if
$V_k=V_k^\text{sp}$. The rest of the conditions are also met: the solution will
satisfy $P_k^\text{min}\le P_k\le P_k^\text{max}$ because the barrier guarantees
that if the starting point is interior, the solution will remain so; and in case
of saturation, the correct control sensitivities are satisfied as well, as it is
readily verified by inspecting~\eqref{eq:DCBarrier}. Therefore this suggests a
HELM scheme where using $\mu=\mu_0(1-s)$ to embed~\eqref{eq:DCBarrier} can
accomplish all the desired goals. The next subsection will fully develop this
idea for the AC case.

\subsection{AC power flow}
This treatment finds an exact parallel in the AC case. However, the authors have
found this is only possible in the limit of an ideal, lossless transmission
network, where all branches (i.e.\ lines, transformers, shunts) have zero
resistance. This is only a limitation as far as it concerns the
\emph{characterization} of the solution provided by the method, not its
precision.  For networks in which this ideality limit is strongly violated, one
can still take the proposed HELM scheme at face value and compute a feasible
power flow satisfying control limits---the only difference is that in this case
it is no longer strictly true that the resulting solution is minimizing
something.  Still, resistances in transmission networks are typically low
($R/X\lesssim 0.1$), so that, simply invoking a continuity argument, one should
expect that using this method should lead to a solution that
\emph{approximately} minimizes the Lagrangian shown here, which will be shown to
be related to the net reactive power losses.

It will first be shown how the power flow equations of an ideal lossless AC
network can be derived from the minimization of a suitable Lagrangian function.
In a lossless network all line and transformer conductances vanish, so that all
admittances, including shunts, become pure imaginary. Using real and imaginary
components, equation~\eqref{eq:PF} becomes:
\begin{equation}
  \label{eq:PF_cartesian}
  \begin{split}
    -b_{ij} \Im(V_i-V_j) - b_i^\text{sh} \Im V_i = \Re I_i + 
    \frac{P_i\Re V_i + Q_i\Im V_i}{|V_i|^2} \\
    b_{ij} \Re(V_i-V_j) + b_i^\text{sh} \Re V_i = \Im I_i + 
    \frac{P_i\Im V_i - Q_i\Re V_i}{|V_i|^2}
  \end{split}
\end{equation}
Written in this form, it is straightforward to verify that these equations can
be derived from the minimization of the following Lagrangian:
\begin{multline}
  \label{eq:ACLagrangian}
  \mathcal{L} = -\frac{1}{2} \sumij b_{ij} \left\{
    \left(\Re V_i-\Re V_j\right)^2 + \left(\Im V_i-\Im V_j\right)^2 \right\} \\
  -\frac{1}{2} \sum_i b_i^\text{sh} \left\{
    \left(\Re V_i\right)^2 + \left(\Im V_i\right)^2 \right\}
  -\frac{1}{2} \sum_i \left( \Re I_i \Im V_i - \Im I_i \Re V_i \right)\\
  -\frac{1}{2} \sum_i Q_i \ln \left\{
    \left(\Re V_i\right)^2 + \left(\Im V_i\right)^2 \right\}
  - \sum_i P_i \tan^{-1} \left( \frac{\Im V_i}{\Re V_i} \right) \;,
\end{multline}
where the index $j$ in the double summation includes the swing bus. If the
complex voltages and complex current injections are now reinterpreted to be
vectors in 2-dimensional real space, \eqref{eq:ACLagrangian} can be written in
the more compact form:
\begin{multline}
  \label{eq:ACLagrangianVec}
  \mathcal{L} = -\frac{1}{2} \sumij b_{ij} \left\| V_i- V_j \right\|^2
  -\frac{1}{2} \sum_i b_i^\text{sh} \left\| V_i \right\|^2 \\
  -\frac{1}{2} \sum_i \left\| I_i \times V_i \right\|
  -\frac{1}{2} \sum_i Q_i \ln \left\| V_i \right\|^2 
  - \sum_i P_i \theta_i  \;,
\end{multline}

Let us analyze the meaning of each term in this expression. The first term is
the sum of the reactive power losses ($I^2X$) over all network branches (divided
by two). Since the susceptance $b_{ij}$ of transmission lines, except in
exceptional cases such as equivalents, is always negative, this first term is
always positive. The next term is similar in nature, as it accounts for the
reactive power injection/consumption due to shunt admittance terms. These
originate mainly from in line charging susceptances, but may also include
contributions from shunt banks and constant-impedance load models.  The next
term, containing the cross product of ``vectors'' $I_i$ and $V_i$ at each bus,
shows the effect of local constant-current injections. Its interpretation in a
mechanical analogue may not be obvious, but note that it also has dimensions of
power. The next term may be interpreted as a sort of potential energy provided
by an external on-site ``field'' produced by a local ``charge'' $Q_i$. It is a
local logarithmic potential acting on each bus. The last term may also be
interpreted as produced by an on-site field, this time mediated by external
charges $P_i$ acting on each node and exerting a sort of transversal force that
tries to rotate the voltage vector. Therefore all terms in this Lagrangian
consist of magnitudes having dimensions of power, and can be interpreted as
being some sort of potential energy, if one forgets about the time dimension.
As in the DC case, one can see this as a pseudo-mechanical analogue, where the
power flow equations are re-interpreted as the equations for the static
equilibria of the system (one could think of particles moving in a 2D space,
interconnected with springs and having on-site potentials acting on them).  This
analogy may be exploited in multiple ways by drawing on the techniques and
intuition from classical mechanics, but here the focus will be on the
constrained problem.

Analogously to the DC case, voltage regulation can be seen as the following
constrained minimization problem:
\begin{equation}
  \label{eq:ACminimiz}
  \begin{split}
    \underset{\{V_i\}}{\text{minimize:}} &\quad \mathcal{E} \left( \{V_i\} \right) \\
    \text{subject to:} &\quad |V_k|^2 - W_k^\text{sp}= 0
    \qquad \left( k\in \{\text{V-regulated}\} \right)
  \end{split}
\end{equation}
Here $\mathcal{E}$ is defined as $\mathcal{L}$ in~\eqref{eq:ACLagrangianVec},
but omitting the terms $\frac{1}{2} Q_k \ln \left\| V_k \right\|^2$ for
voltage-regulated buses $k$. Again, this problem is solved with the standard
technique of Lagrange multipliers:
\begin{equation}
  \label{eq:ACLagrangian2}
  \mathcal{L} \equiv \mathcal{E}
  \; - \frac{1}{2} \sum_k Q_k \ln \frac{|V_k|^2}{W_k^\text{sp}}  \;,
\end{equation}
which is the same as~\eqref{eq:ACLagrangianVec}, except for additional constant
terms $Q_k\ln W_k^\text{sp}$ which do not depend on the voltages. The
stationarity conditions then take the same form as the original power flow
equations~\eqref{eq:PF_cartesian}. Similarly to the DC case, reactive power
injections can be seen as the Lagrange multipliers of minimization
problem~\eqref{eq:ACminimiz}.  The physical interpretation for the minimization
of $\mathcal{E}$ in terms of energy is in general not so evident as in DC, but
it can still be recovered if all active power injections $P_i$ are made zero. In
that limit, it can be shown that all angles are zero, all voltages and currents
are real, and the system actually becomes mathematically equivalent to the DC
case.

% In this sense, it may be argued that, at least in transmission networks, the
% reactive power subproblem constitutes the foundational backbone for the full
% AC problem.

Let us now contemplate resource constraints, i.e., Mvar limits:
\begin{equation}
  \label{eq:ACminLimited}
  \begin{split}
    \underset{\{V_i\}}{\text{minimize:}} &\quad \mathcal{E} \left( \{V_i\} \right) \\
    \text{subject to:} &\quad |V_k|^2 - W_k^\text{sp}= 0
    \qquad \left( k\in \{\text{V-regulated}\} \right) \\
    &\quad Q_k^\text{min} \le Q_k \le Q_k^\text{max}
  \end{split}
\end{equation}
These inequality constraints are on the dual variables of
problem~\eqref{eq:ACminimiz}. Consider then the maximization of the dual
problem by means of logarithmic barrier methods, by defining:
\begin{multline}
  \label{eq:ACBarrier}
  \mathcal{B}_\mu \left( \{Q_k\} \right) \equiv \mathcal{E}
   - \frac{1}{2} \!\sum_k Q_k \ln \frac{|V_k|^2}{W_k^\text{sp}} \\
  + \mu \ln \left(Q_k^\text{max}-Q_k\right)
  + \mu \ln \left(Q_k-Q_k^\text{min}\right)  
\end{multline}
where $\mu>0$ is the barrier parameter. The
stationarity conditions for maximizing $\mathcal{B}_\mu$ yield:
\begin{equation}
  \label{eq:ACBarrierCond1}
  - \frac{1}{2} \ln \frac{|V_k|^2}{W_k^\text{sp}}
  - \frac{\mu}{Q_k^\text{max}-Q_k} + \frac{\mu}{Q_k-Q_k^\text{min}} = 0
\end{equation}
Rearranging, one obtains:
\begin{multline}
  \label{eq:ACBarrierCond2}
  \left(\ln |V_k|^2 - \ln W_k^\text{sp} \right)
  \left(Q_k^\text{max}-Q_k\right) \left(Q_k-Q_k^\text{min}\right) = \\
  2 \mu \left( Q_k^\text{max} + Q_k^\text{min} - 2Q_k \right)
\end{multline}
In the limit $\mu\rightarrow0$, this equation is essentially the same as the
complementarity constraint~\eqref{eq:complementarity}, since
$\ln |V_k|^2 = \ln W_k^\text{sp}$ if and only if $|V_k|^2 = W_k^\text{sp}$.  The
rest of the conditions are also met: \eqref{eq:Qlims} is satisfied because the
barrier terms guarantee that the solution will remain interior if the starting
point is so; and the signs of the control errors, given
by~\eqref{eq:sensitiv_cond}, can be verified to be the correct ones by
inspecting~\eqref{eq:ACBarrierCond1}.  When the solution is such that the
$Q_k^\text{max}$ constraint is active, one obtains $|V_k|^2 \le W_k^\text{sp}$;
and vice versa.

Therefore this suggests a HELM scheme where~\eqref{eq:ACBarrierCond1} is
embedded by using, for instance, $\mu=\mu_0(1-s)$. Together with some additional
precautions, to be described in the next section, this equation
replaces~\eqref{eq:PVulimconstraint} and provides a HELM scheme to solve the
problem of Q-limits.  This establishes the relationship between the proposed
HELM scheme and the above minimization problem. The HELM solution is thus
characterized as minimizing $\mathcal{E}$ in~\eqref{eq:ACminLimited}, which, in
the limit of the ideal lossless network, is essentially the net amount of
reactive losses.

Finally, this result also suggests an apparent resemblance between so-called
Interior Point Methods (IPM) and the analytic continuation technique used in
HELM. However, just as in the case of homotopy methods, IPM and similar barrier
methods are based on continuity and differentiability, not holomorphicity (which
is a much stronger requirement).

\section{HELM scheme. Consistency requirements}
\label{sec:HELMscheme}

Following the above, a HELM scheme is proposed to solve the problem of voltage
regulation while observing Q-limits, using the canonically embedded
system~\eqref{eq:PF_PQ}--\eqref{eq:PF_PV} and
embedding~\eqref{eq:ACBarrierCond1} as follows:
\begin{equation}
  \label{eq:compl_embed}
  V_k(s) V_k^*(s^*) - W_k^\text{sp} = \frac{\mu_k(1-s)}{Q_k(s)-Q_k^\text{min}}
                                  - \frac{\mu_k(1-s)}{Q_k^\text{max}-Q_k(s)}
\end{equation}
where $\mu_k$ are suitable constants to be discussed below (following IPM
methods, they are chosen to be the same for both terms). As discussed in the
previous section, in the limit $s\rightarrow0$ these equations reproduce the
complementarity
constraints~\eqref{eq:complementarity}--\eqref{eq:sensitiv_cond}. At this point,
it is important to remark that there are possibly many other methods to obtain
solutions satisfying the complementarity equations, but this is a method
\emph{informed by the Lagrangian}. In case there are several feasible solutions,
the method presented here will select the one minimizing the
Lagrangian~\eqref{eq:ACLagrangianVec}, which has shown to be rooted in the
physics of the transmission network.

On the other hand, the following treatment will consider the possibility that
the voltage regulation may be \emph{remote}, i.e.\ that the controlled bus and
its controlling injection are not on the same bus. There is nothing in the
constraint equations preventing this, as it would still produce the same,
well-defined algebraic problem. The only difference in such case is that, in a
strict sense, it would no longer be possible to interpret the resulting solution
as the minimizer of the constrained problem shown in the previous
section. However, if one restricts this to the most common case, which is
regulation at a distance of one bus, the interpretation is expected to still
hold approximately, in the sense that other feasible solutions would yield a
higher value of the Lagrangian.

Therefore from now on the notation will label the set of $\Npv$ reactive bus
injections with indexes $a$, and the set of their corresponding $\Npv$ regulated
bus voltages with $b$. Note that the treatment given here does not contemplate
concurrent control (i.e. shared responsibility regulation) of the same bus
originated from different buses. This problem has no unique solution in general,
as it depends on the details of the var-sharing policy, priority rules among
generators, etc.

% Note: about needing extra info about controls when there's concurrency:
% * The paper~\cite{Peterson71} contains an interesting discussion about the
%   priority of concurrent control devices, but the method does not enforce
%   limits.
% * Chang and Brandwajn~\cite{ChangBrandwajn88} also contains interesting
%   discussions at the end about priorities of concurrent controls, multiplicity
%   of solutions, etc.

HELM now requires that the embedding is designed in a way that a unique and
meaningful reference powerflow state is obtained at $s=0$: the zero-injection
state, in which all voltages are equal to the swing. It is then postulated that
the \emph{operational} solution, in contrast to the many other power flow
solutions, is the one that is analytically continued to $s=1$ from this
reference state (when it exists). This is based on physical arguments, discussed
at length in~\cite{Trias12,Trias15}.  An important question then is
whether~\eqref{eq:compl_embed} can be consistent with the HELM reference
solution, given that limits $Q^{min},Q^{max}$ come from engineering constraints
and could have any value. This consistency can be examined by considering the
power series of the variables involved, and checking ~\eqref{eq:compl_embed} at
order zero. Using the notation $V_b(s)=\sum_nV_b[n]s^n, Q_a(s)=\sum_nQ_a[n]s^n$
for power series coefficients, one obtains at zero order:
\begin{equation*}
  V_b[0] V_b^*[0] - W_b^\text{sp} = \frac{\mu_a}{Q_a[0]-Q_a^\text{min}}
                                - \frac{\mu_a}{Q_a^\text{max}-Q_a[0]}
\end{equation*}
and since the reference solution has $V_b[0]=1,\, Q_a[0]=0$ for all buses:
\begin{equation}
  \label{eq:zero_order}
  1 - W_b^\text{sp} = 
  -\mu_a \frac{Q_a^\text{max}+Q_a^\text{min}}{Q_a^\text{max}Q_a^\text{min}}
\end{equation}

There are two different consistency requirements to observe here. The first one
is that the reference state should not violate limits (or hit exactly a limit),
which implies $Q_a^\text{min}<0, Q_a^\text{max}>0$.  In the parlance of interior
point methods, this is is the analogue of ensuring that the starting point is
``interior''. The second requirement is that~\eqref{eq:zero_order} should be
satisfied with a \emph{positive} value of the barrier constant $\mu_a$. Since
$Q_a^\text{min}Q_a^\text{max}<0$, this implies
$\sgn(1-W_b^\text{sp})=\sgn(Q_a^\text{max}+Q_a^\text{min})$. Neither one of
these two requirements are guaranteed to be satisfied, in general. Therefore one
is forced to embed both the limits and the setpoints, in a way that the above
conditions are satisfied at order zero. Note how this was already done in the
unlimited PV case, as $W_b^\text{sp}(s)=1+s(W_b^\text{sp}-1)$
in~\eqref{eq:PVulimconstraint}. In this case, the embeddings should be designed
to require that:
\begin{equation}
  \begin{split}
    & Q_a^\text{min}[0]<0 ,\quad Q_a^\text{max}[0]>0 \\
    & \sgn\left(1-W_b^\text{sp}[0]\right) = 
    \sgn\left(Q_a^\text{max}[0]+Q_a^\text{min}[0]\right)
  \end{split}
\end{equation}
For one-sided limits, the last expression needs to be replaced by
$W_b^\text{sp}[0]>1$ when only $Q_a^\text{max}$ is present, or by
$W_b^\text{sp}[0]<1$ when only $Q_a^\text{min}$ is present.  All these
conditions can be accomplished by means of many possible embeddings. However, in
keeping with the general principle of introducing the minimum amount of changes
needed, our preferred method is to embed limits and setpoints linearly, choosing
suitable guard constants:
\begin{equation}
  \label{eq:Qlims_embed}
  \begin{split}
    Q_a^\text{min}(s) &= Q_a^\text{min} + \delta Q_a^\text{min} (1-s) \\
    Q_a^\text{max}(s) &= Q_a^\text{max} + \delta Q_a^\text{max} (1-s) \\
    W_b^\text{sp}(s)  &= W_b^\text{sp}  + \delta W_b^\text{sp} (1-s)
  \end{split}
\end{equation}
If both limits are present, one reasonable choice for these constants consists
in fixing $\mu_a=1$ and using the same embedding
for $W_b^\text{sp}$ as in the unlimited case, plus the following symmetrizing
embedding for the limits:
\begin{equation}
  \label{eq:JLMembedding}
  \begin{split}
    Q_a^\text{min}(s) &= - \frac{Q_a^\text{max}-Q_a^\text{min}}{2}
     + s \frac{Q_a^\text{max}+Q_a^\text{min}}{2} \\
    Q_a^\text{max}(s) &= \frac{Q_a^\text{max}-Q_a^\text{min}}{2}
     + s \frac{Q_a^\text{max}+Q_a^\text{min}}{2}
  \end{split}
\end{equation}
which leads to the following embedding constants in~\eqref{eq:Qlims_embed}:
\begin{equation}
  \label{eq:JLMembeddingconstants}
  \begin{split}
    \delta Q_a^\text{min} &= \delta Q_a^\text{max} = - \frac{Q_a^\text{max}+Q_a^\text{min}}{2} \\
    \delta W_b^\text{sp}  &= 1 - W_b^\text{sp}
  \end{split}
\end{equation}
For buses with one-sided limits, one needs to choose values for the guard
constants in~\eqref{eq:Qlims_embed} so that the zero order coefficients satisfy
the sign requirements and have reasonable values.  For instance, in case one has
only a lower limit $Q_a^\text{min}$:
\begin{equation}
  \label{eq:JLMembeddingconstants1limmin}
  \begin{split}
    \delta Q_a^\text{min} &= Q_a^\text{min}[0] - Q_a^\text{min} \\
    \delta W_b^\text{sp}   &= W_b^\text{sp}[0] - W_b^\text{sp} \\
    \mu_a &= Q_a^\text{min}[0] \left( W_b^\text{sp}[0] - 1 \right)
  \end{split}
\end{equation}
so that one reasonable choice could be $W_b^\text{sp}[0]=0.9$ pu and
$Q_a^\text{min}[0]=-10$ pu.  For buses where there is only an upper limit
$Q_a^\text{max}$, one obtains:
\begin{equation}
  \label{eq:JLMembeddingconstants1limmax}
  \begin{split}
    \delta Q_a^\text{max} &= Q_a^\text{max}[0] - Q_a^\text{max} \\
    \delta W_b^\text{sp}  &= W_b^\text{sp}[0] - W_b^\text{sp} \\
    \mu_a &= Q_a^\text{max}[0] \left( W_b^\text{sp}[0] - 1 \right)
  \end{split}
\end{equation}
and the corresponding reasonable choice could be $W_b^\text{sp}[0]=1.1$ pu and
$Q_a^\text{min}[0]=+10$ pu.  However, to keep the following treatment as general
as possible, no particular embedding constants will be assumed; it will only be
assumed that the embeddings are linear, as given by~\eqref{eq:Qlims_embed}.

The HELM procedure can now be constructed by obtaining the linear system that
allows obtaining all unknown coefficients at order $N$ from the knowledge of
coefficients at orders $N-1$ and lower. This system has been referred to as the
$N$-th order representation of the embeded equations~\cite{Trias12}. Solving
these linear systems in sequence is what constructs the germ of the reference
solution, which is later to be analytically continued into the final solution at
$s=1$ (if it exists). For the base power flow equations~\eqref{eq:PF_PQ} and
\eqref{eq:PF_PV}, \ref{app:methsummary} provides the $N$-th order representation
formulas (in the more general Pad\'e-Weierstrass framework). Here the focus is
put on the constraint equations, in order to show how their contribution results
in an enlarged but well-defined linear system.

The embedding selected in~\eqref{eq:Qlims_embed} directly provides the power
series corresponding to setpoints and limits. They have just two coefficients:
\begin{equation*}
  \begin{split}
    Q_a^\text{min}(s) &= Q_a^\text{min}[0] + Q_a^\text{min}[1] \, s \\
    Q_a^\text{max}(s) &= Q_a^\text{max}[0] + Q_a^\text{max}[1] \, s \\
    W_b^\text{sp}(s)  &= W_b^\text{sp}[0]  + W_b^\text{sp}[1] \, s    
  \end{split}
\end{equation*}
If you chose the symmetrizing embedding~\eqref{eq:JLMembedding}, the values of
$\mu_a$ can be chosen all $\mu_a=1$; otherwise they will be given by the
constraint equation~\eqref{eq:compl_embed} at zero order:
\begin{equation*}
  \mu_a = \left( W_b^\text{sp}[0] - 1 \right)
  \frac{Q_a^\text{max}[0]\,Q_a^\text{min}[0]}{Q_a^\text{max}[0]+Q_a^\text{min}[0]}
\end{equation*}
Now the following auxiliary series will be defined:
\begin{equation*}
  \mathcal{B}_a^{(+/-)}(s) \equiv \frac{\mu_a(1-s)}{Q_a(s)-Q_a^\text{(min/max)}(s)} =
  \sum_{m=0}^\infty \mathcal{B}_a^{(+/-)}[m] \, s^m
\end{equation*}
Since $Q_a[0]=0$, its zero-order values are:
\begin{equation*}
  \mathcal{B}_a^{(+/-)}[0] = -\frac{\mu_a}{Q_a^\text{(min/max)}[0]}
\end{equation*}

For $N\ge 1$, equating the $N$-th power coefficients on each side of the constraint
equation~\eqref{eq:compl_embed} yields:
\begin{equation}
  \label{eq:Nth_constraint}
  \sum_{m=0}^N V_b[m]\,V_b^*[N-m] - W_b^\text{sp}[1]\,\delta_{1,N} =
  \mathcal{B}_a^-[N] + \mathcal{B}_a^+[N]
\end{equation}
where $\delta_{1,N}$ is the Kronecker delta. Using the known values $V_b[0]=1$,
the convolution sum on the right hand side decomposes as:
\begin{equation*}
  2\Re V_b[N] + \sum_{m=1}^{N-1} V_b[m]\,V_b^*[N-m]
\end{equation*}
Note that, even though this expression involves complex magnitudes, the sum is
real, given the symmetry of the convolution.

The power series coefficients of $\mathcal{B}_a^{(+/-)}$ can be obtained in
terms of those of $Q_a(s)$ from their definition. For instance, by equating the
$N$-th order coefficients of each side of equation
$\mathcal{B}_a^+(s)\left(Q_a(s)-Q_a^\text{max}(s)\right) = \mu_a(1-s)$, for
$N\ge 1$:
\begin{multline*}
  \sum_{m=0}^N \mathcal{B}_a^+[m] \, Q_a[N-m]
  - Q_a^\text{max}[0] \, \mathcal{B}_a^+[N] \\
  - Q_a^\text{max}[1] \, \mathcal{B}_a^+[N-1] = -\mu_a\,\delta_{1,N}
\end{multline*}
and analogously for $\mathcal{B}_a^-$. Since $Q_a[0]=0$, one obtains:
\begin{multline*}
  Q_a^\text{max}[0] \, \mathcal{B}_a^+[N] = \mathcal{B}_a^+[0]\, Q_a[N]
  + \sum_{m=1}^{N-1} \mathcal{B}_a^+[m] \, Q_a[N-m] \\
  - \mathcal{B}_a^+[N-1] \, Q_a^\text{max}[1] + \mu_a\,\delta_{1,N}
\end{multline*}
Moving all terms in~\eqref{eq:Nth_constraint} that include coefficients at order
$N$ to the left hand side, one obtains:
\begin{equation}
  \label{eq:Nth_constraint2}
  2\Re V_b[N] - \left( \frac{\mathcal{B}_a^-[0]}{Q_a^\text{min}[0]}
              + \frac{\mathcal{B}_a^+[0]}{Q_a^\text{max}[0]} \right) Q_a[N] =
  \mathcal{T}_a[N-1]
\end{equation}
where the symbol $\mathcal{T}_k[N-1]$ is used as a convenient shorthand to group
all other terms, which are either known quantities or involve coefficients of
order $N-1$ and lower. In the case of one-sided limits, the expressions are very
similar and straightforward to obtain.

This set of equations~\eqref{eq:Nth_constraint2} complement the rest, to form a
linear system of dimension $2\Npq+3\Npv$, involving variables $\Re V_j[N]$,
$\Im V_j[N]$, and $Q_a[N]$. Actually, variables $Q_a[N]$ can be easily
eliminated from~\eqref{eq:Nth_constraint2} in terms of $\Re V_b[N]$ (it is
verified that their pivot cannot be zero), and substituted into the rest of the
equations to obtain a linear system of dimension $2(\Npq+\Npv)$. The method then
proceeds as usual~\cite{Trias12}: one solves these linear systems in sequence,
starting from $N=1$ until enough power series terms are obtained to either
obtain convergence of detect oscillation in the Padé approximant
sequences.

\section{The Padé-Weierstrass method for analytic continuation}
\label{sec:PW}

The preceding scheme is able to produce a well-defined germ, but its analytic
continuation to $s=1$ by means of Padé approximants suffers from an intrinsic
problem: by its very construction, the embedded complementarity
constraint~\eqref{eq:compl_embed} creates a singularity of the algebraic curve
at $s=1$, since this is a branching point for many other solutions. Such
solutions correspond to the many possible choices in which one can saturate the
controls (i.e.\ different selections of buses switched to PQ-type). All of them
satisfy the power flow equations as well as the complementarity constraint, but
most of them violate one or more of the other conditions, i.e.\ \eqref{eq:Qlims}
and \eqref{eq:sensitiv_cond}. And if there existed any configuration satisfying
all the requirements (something that cannot be ruled out, due to the
nonlinearity of the problem), it would be ``worse'' than the solution produced
by the proposed HELM scheme, in the sense of the minimization criterion provided
by the Lagrangian formulation on which the method is grounded (the HELM solution
approximately minimizes reactive power losses).

As it is well-known from Stahl's theory~\cite{Stahl97}, the near-diagonal
sequences of Padé approximants cease to converge at singularities (and on the
minimal cut-set joining them). For the HELM scheme developed in the previous
section this means that, even if the power flow is feasible, the approximants
are no longer guaranteed to converge at $s=1$. Actually, unless the case is
outright infeasible (in which case the approximants will clearly oscillate),
what happens is that the convergence rate gets very slow as $s$ approaches 1.
In the unlimited HELM method, this would only happen when calculating a case
\emph{exactly} at a point of collapse. In both situations, due to the limits of
floating point arithmetic (truncation and round-off), there is a point at which
high order Padé approximants can no longer be calculated with enough precision
to improve the results. The net result is a ceiling to the achievable precision
in the solution as $s\rightarrow 1$. Numerical experiments show that, in the
particular case of the Q-limits HELM scheme, the magnitude of the precision loss
is usually too large to obtain acceptable results.  On the other hand, in case
of powerflow infeasibility, a voltage collapse point will be met earlier at some
$s_\text{crit}<1$, as shown in~\cite{Trias15}; this will be more easily detected
as oscillations in the approximants when evaluated at $s=1$.

In any case, given the nature of the HELM reference solution at $s=0$, it is
always possible to find a real value $0<s_0<1$ such that the Padé approximants
do converge to the required precision. This means it is always possible to
obtain a \emph{partial solution} $V_i(s_0)$ whose values are exact, within
machine precision. What follows is a method that exploits this property and
comes up with a method to perform the analytic continuation of the reference
germ in a series of stages that approach $s\rightarrow 1$, greatly enhancing the
numerical precision. This is effectively an analytic continuation along a path
on the real axis, in the same spirit as Weierstrass, but using a novel
technique. A classical Weierstrass procedure would use the initial power
series to construct a second power series at a point $s_0$ inside its radius of
convergence, and then repeat this process.  However,
Henrici~\cite{Henrici93Vol1} shows that this naive approach yields in practice
very poor numerical precision. Here, by contrast, the method uses partial
solutions and the power flow equations in order to re-expand the new germs along
the path.  Since the procedure makes use of Padé approximants as a key element
for obtaining partial solutions to very high precision, the method is termed
\emph{Padé-Weierstrass} (to be abbreviated as P-W).

\subsection{Two-bus DC system}
The essential elements of the Padé-Weierstrass method can be shown in the
simplest power flow problem, the two-bus DC system. For further simplicity, the
swing voltage will be normalized to 1. The embedded equation reads:
\begin{equation}
  \label{eq:2busDC}
  G\Bigl( V(s) - 1 \Bigr) = s\frac{P}{V(s)}
\end{equation}
As always, the aim is to obtain the analytic continuation of the HELM germ to
$s=1$. Assume now that the convergence of Padé approximants is so slow at $s=1$
that the achievable precision is degraded (though in this simple system, this
only happens when the solution is \emph{very} close to point of voltage
collapse). One can always find a value $0<s_0<1$ such that the approximants
converge to any required tolerance (within machine precision limits). Let us use
these values $s_0$ and $V(s_0)$ to rewrite the equation in terms of the
following renormalized parameter and voltage:
\begin{equation}
  \label{eq:renorm}
  \begin{split}
    s &\equiv s_0 + (1-s_0)s' \\
    V(s) &\equiv V(s_0) V'(s')
  \end{split}
\end{equation}
The parameter change $s\rightarrow s'$ is a conformal mapping that takes the
range $[s_0,1]$ in $s$-space into the range $[0,1]$ in $s'$-space. Using these,
\eqref{eq:2busDC} reads:
\begin{equation*}
  V(s_0) \, G \Bigl( V(s_0)V'(s') - 1 \Bigr) =
  \frac{s_0 P}{V(s)} + s' \frac{(1-s_0)P}{V'(s')}
\end{equation*}
Using the fact that $G\bigl(V(s_0)-1\bigr)=s_0\frac{P}{V(s_0)}$, the right hand
side of this equation yields:
\begin{multline*}
  V(s_0)\,G\Bigl(V(s_0)-1\Bigr) V'(s') + V(s_0)\,G\Bigl(V'(s')-1\Bigr) = \\
  s_0 P\,V'(s') + V(s_0)\,G\Bigl(V'(s')-1\Bigr) 
\end{multline*}
and the final equation becomes:
\begin{equation}
  \label{eq:PWstep}
  G' \Bigl( V'(s') - 1 \Bigr) = \Gamma' \Bigl( \frac{1}{V'(s')} - V'(s') \Bigr) 
  + s' \frac{P'}{V'(s')}
\end{equation}
where the new constants are defined as $G'\equiv V(s_0)G$,
$\Gamma'\equiv s_0 P$, and $P'\equiv(1-s_0)P$. This equation describes a new
power flow problem, very similar to~\eqref{eq:2busDC} except for the appearance
of two new terms which will be jointly referred to as a $\Gamma$-term. Before
analyzing in detail its meaning and role, let us remark that the term itself is
completely invariant under this renormalization procedure. That is, assume that
instead of~\eqref{eq:2busDC} one is given the following initial power flow problem:
\begin{equation}
  \label{eq:2busDC_withGamma}
  G\Bigl( V(s) - 1 \Bigr) = \Gamma \Bigl( \frac{1}{V(s)} - V(s) \Bigr)
  + s\frac{P}{V(s)}
\end{equation}
Then it can be verified that, under the change~\eqref{eq:renorm}, this system
also transforms exactly into system~\eqref{eq:PWstep}, with these new
parameters:
\begin{equation*}
  \begin{split}
    G' &\equiv V(s_0)G \\
    \Gamma' &\equiv \Gamma + s_0 P \\
    P' &\equiv (1-s_0)P
  \end{split}
\end{equation*}

This result is quite remarkable. Conceptually, this is similar to the way
certain changes of coordinates leave physical laws invariant, i.e. the principle
of general covariance.  For instance Newton's equations of classical mechanics
are invariant under Galilean transformations, and Einstein's equations of
Special Relativity are invariant under Lorentz transformations. What we have
here is that, when the embedded powerflow problem is written
as~\eqref{eq:2busDC_withGamma}, it is invariant under the ``change of reference
frame'' given by~\eqref{eq:renorm}.

This property can be exploited to good effect. The whole procedure described
here, which will be referred to as a single \emph{Padé-Weierstrass stage}, can
be repeated multiple times, for suitable values $s'_0, s''_0$, etc.  At each
stage one always obtains a new powerflow problem that it is formally the same,
where only the parameters have changed. The great advantage is that, as it will
be shown, each successive stage produces a better-conditioned HELM problem from
the numerical point of view. Then, if at some stage $k$ the Padé approximants
are able to converge with the requested precision at $s^{(k)}=1$, one can simply
undo all the changes~\eqref{eq:renorm} and thus obtain the sought solution to
the initial problem.

Let us now turn to the $\Gamma$-term. On the first P-W stage it is zero, but it
will appear at all subsequent stages. The term can be thought of as a
combination of a constant-power injection and a constant-admittance shunt. Their
respective currents cancel out when $V'(s')=1$, which happens at $s'=0$.  This
is consistent with the HELM reference state (no flows, all voltages equal to the
swing). However, the novelty is that this $\Gamma$-term is not embedded in $s$,
so it remains to be seen how the HELM procedure can work. It turns out that
given the specific form of this term, whereby their currents cancel out at
$s=0$, the standard HELM method can be carried through without any problem, as
shown now.

To kickstart the method on~\eqref{eq:2busDC_withGamma}, consider first the
equation obtained for the power series coefficients at zero order. It is
verified to be consistent with the HELM reference state, $V(0)=1$, since
$V^{-1}[0]=V[0]=1$. Recall that the power series coefficients for the function
$1/V(s)$, symbolically represented as $V^{-1}[N]$ (not to be confused with
$1/V[N]$), can be obtained from those of $V(s)$ by the relationship:
\begin{equation}
  \label{eq:invSeries}
  \sum_{m=0}^N V[m] \, V^{-1}[N-m] = \delta_{N,0}
\end{equation}
At orders $N\ge 1$, one obtains:
\begin{equation*}
  G V[N] = \Gamma \Bigl( V^{-1}[N] - V[N] \Bigr) + P V^{-1}[N-1]
\end{equation*}
Using~\eqref{eq:invSeries}, one finally obtains:
\begin{equation}
  \label{eq:2busDC_Norder}
  \left( G + 2\Gamma \right) V[N] =
  P V^{-1}[N-1] - \Gamma\sum_{m=1}^{N-1} V[m] \, V^{-1}[N-m]
\end{equation}
All terms on the right hand side are order $N-1$ or lower, so they can be
computed from coefficients obtained at previous steps. Therefore the HELM scheme
can be carried out without any problem.

There is one final aspect that deserves attention. The coefficient of $V[N]$
in~\eqref{eq:2busDC_Norder} could in principle become zero if left unchecked. In
a full $n$-bus system, the analogous problem would manifest itself as the matrix
of the HELM linear system becoming singular. Let us analyze for instance what
would happen when trying to solve the two-bus system~\eqref{eq:2busDC} for a
case right at the voltage collapse point, i.e.\
$P=P_\text{crit}=-G/4$. Performing one step of the P-W procedure, one chooses a
point $0<s_0<1$ for which the approximants converge to $V(s_0)$ within the
required precision. Since this problem can be solved in closed form, one finds
$V(s_0)=(1+\sqrt{1-s_0})/2$. Therefore, at the next P-W stage the pivot
coefficient in~\eqref{eq:2busDC_Norder} becomes:
\begin{equation*}
  G' + 2\Gamma' = \frac{1}{2} G \left( 1-s_0 + \sqrt{1-s_0} \, \right)
\end{equation*}
From this expression one can observe that, when the target state at $s=1$ sits
near or at a singularity, there is a certain trade-off in the choice of
$s_0$. Even if the approximants allowed it, getting too close in one step
(i.e. $s_0\approx 1$) would produce a numerically unstable system in the next
P-W step; so it may be more advantageous to settle for a smaller value even if
this means taking a few more P-W steps, if the overall numerical stability
improves.

Next it will be shown how this method can be extended to the full AC power flow,
including voltage regulation and Q-limits.

\subsection{Full AC system: main equations}

The P-W method will now be developed for the full AC power flow problem, as
represented by the embedded system~\eqref{eq:PF_PQ}--\eqref{eq:PF_PV} and
including the observance of Q-limits by means of the complementarity
constraint~\eqref{eq:compl_embed} as shown in Section~\ref{sec:HELMscheme}. As
in the two-bus DC system, the end result will be a sequence of HELM subproblems
all formally identical to the initial one, where only the parameters get
re-defined. At each successive stage of the P-W process, the numerical stability
of each HELM subproblem improves, so that it is possible to reach much higher
levels of precision at $s=1$, even when a singularity is very close. When
dealing with limits, this is a must, since the constraint equation creates a
singularity exactly at $s=1$, as discussed at the beginning of this Section.

In order to simplify the exposition, constant-current injections (either load or
generation) will be omitted for now; their treatment is not given here but, it
can be shown that the method is easily adapted to contemplate them as well. Let
us first consider the equation for PQ buses, \eqref{eq:PF_PQ}, which here will
be written as: \begin{equation}
  \label{eq:PW_PQ}
    i \in \text{PQ:}  \; \sum_{j=0}^n Y_{ij} V_j(s) + sY_i^\text{sh} V_i(s) =
    \frac{sS_i^*}{V_i^*(s^*)} + \Gamma_i \left(\frac{1}{V_i^*(s^*)} - V_i(s)\right)
\end{equation}
Here $\Gamma_i$ is a parameter whose value is zero on the first stage of the
method, but in subsequent stages it will be non-null, in general. As in the DC
example, assume now that the system is solved with the standard HELM method and
a real value $0 < s0 <1$ is chosen such that the Padé approximants converge with
the required tolerance. The values $s_0$ and $V_i(s_0)$ will now be used to
rewrite the equation in terms of the renormalized parameter $s'$ and voltage
variables $V'(s')$ defined by~\eqref{eq:renorm}.  Evaluating~\eqref{eq:PW_PQ} at
$s=s_0$ and multiplying both sides by $V_i^*(s_0^*)$, one has:
\begin{equation}
  \label{eq:PW_PQ_at_s0}
  \sum_{j=0}^n V^*_i(s_0) Y_{ij} V_j(s_0) + s_0 Y_i^\text{sh} |V_i(s_0)|^2 =
  s_0 S_i^* +
  \Gamma_i \left( 1 - |V_i(s_0)|^2 \right)
\end{equation}
where the fact that $s_0$ is chosen real has been used to simplify the
expression. A non-real value of $s_0$ could in principle be chosen instead, thus
leading to a path analytic continuation that would meander around the complex
plane in order to get from $s=0$ to $s=1$. Uniqueness of the result would still
be guaranteed, since by Stahl's theorem Padé approximants will keep avoiding the
minimal cut set, and therefore there will be no ``branch jumps'' no matter what
path is chosen.  However, as shown in~\cite{Trias15}, physics-based arguments
dictate that analytic continuation of the white germ is actually \emph{required}
to exist along all points $0\le s_0\le 0$ on the real axis, in order to be
called the operational solution. Therefore the method requires that $s_0$ is
chosen real.

Writing~\eqref{eq:PW_PQ} in terms of~\eqref{eq:renorm}, also multiplying both
sides by $V_i^*(s_0^*)$:
\begin{multline}
  \label{eq:PW_PQ_tmp}
  \sum_{j=0}^n V^*_i(s_0) Y_{ij} V_j(s_0) \, V'_j(s') \\
  + s_0 |V_i(s_0)|^2 Y_i^\text{sh} \, V'_i(s')
  + s' (1-s_0) |V_i(s_0)|^2 Y_i^\text{sh} \, V'_i(s') = \\
  \frac{s_0 S_i^*}{V_i'^*(s'^*)} + s'\frac{(1-s_0) S_i^*}{V_i'^* (s'^*)}
  + \Gamma_i \left(\frac{1}{V_i'^*(s'^*)} - |V_i(s_0)|^2 V_i(s')\right)
\end{multline}
In passing, note that if the swing voltage (index $j=0$) is embedded as
$V_0=1+s(V_\text{sw}-1)$, then it is straightforward to obtain its expression in
terms of the new parameter:
\begin{equation*}
  V_0'(s') = \frac{V_0(s)}{V_0(s_0)} =
  1 + s' \frac{(1-s_0)(V_\text{sw}-1)}{1+s_0(V_\text{sw}-1)} 
\end{equation*}

It is useful to introduce now the shorthand notation:
\begin{equation*}
  \hat{Y}_{ij} \equiv  V^*_i(s_0) Y_{ij} V_j(s_0)
\end{equation*}
However, unlike the original matrix $Y_{ij}$, this new admittance matrix would not
satisfy the transmission condition (i.e., the sum of all columns does not yield
the zero column vector). In fact, the sum of its columns can be readily
calculated using~\eqref{eq:PW_PQ_at_s0}:
\begin{equation*}
  \sum_{j=0}^n \hat{Y}_{ij} = - s_0 Y_i^\text{sh} |V_i(s_0)|^2 +
  s_0 S_i^* +  \Gamma_i \left( 1 - |V_i(s_0)|^2 \right)  
\end{equation*}
This can be fixed by defining the new admittance matrix as follows:
\begin{equation}
  \label{eq:Yrenorm}
  Y'_{ij} \equiv
  \begin{cases}
    \hat{Y}_{ij}                           & \text{if } i \ne j \\
    \hat{Y}_{ii} - \sum_{l=0}^n \hat{Y}_{il}  & \text{if } i=j 
  \end{cases}
\end{equation}
With this definitions and after straightforward algebraic rearrangements and
simplifications of~\eqref{eq:PW_PQ_tmp}, one arrives at the following equation
for PQ buses:
\begin{equation}
  \label{eq:PW_PQ_at_sprime}
  \sum_{j=0}^n Y'_{ij} V'_j(s') + s' Y_i^{\prime\text{sh}} V'_i(s') =
  \frac{s'S_i'^*}{V_i'^*(s'^*)}
  + \Gamma'_i \left(\frac{1}{V_i'^*(s'^*)} - V'_i(s')\right)
\end{equation}
where the new parameters are given by:
\begin{equation*}
  \begin{split}
    Y_i^{\prime\text{sh}} &\equiv (1-s_0) |V_i(s_0)|^2 Y_i^\text{sh}  \\
    \Gamma'_i &\equiv \Gamma_i + s_0 S_i^* \\
    S_i' &\equiv (1-s_0) S_i
  \end{split}
\end{equation*}
Equation~\eqref{eq:PW_PQ_at_sprime} is exactly the same as the original one,
\eqref{eq:PW_PQ}. In this sense, it can be said that the system is invariant
under the Padé-Weierstrass transformation given by~\eqref{eq:renorm}.

Analogously, the equations for PV buses can be shown to be invariant as
well. Following the notation introduced in Section~\ref{sec:HELMscheme}, which
allows for one-on-one remote voltage regulation, the equations corresponding to
\emph{regulating} buses are written as:
\begin{multline}
  \label{eq:PW_PV}
    a \in \text{PV:}  \\ \sum_{j=0}^n Y_{aj} V_j(s) + sY_a^\text{sh} V_a(s) =
    \frac{sP_a - jQ_a(s)}{V_a^*(s^*)} +
    \Gamma_a \left(\frac{1}{V_a^*(s^*)} - V_a(s)\right)
\end{multline}
Note that these buses are not strictly speaking PV-type, as the formulation here
is more general. Buses labeled $a$ control the voltage of buses labeled $b$; but
a bus is said to be PV-type when $a=b$. However, for the sake of simplicity, the
set of buses labeled $a$ will be referred to as ``PV buses'' just to avoid
longer expressions such as ``the buses having voltage-controlling reactive
injections''.

In this case, one also needs to specify how the new variables $Q_a(s)$ should
transform.  In contrast to voltages, it turns out that the transformation that
works is \emph{additive}:
\begin{equation}
  \label{eq:renormQ}
    Q_a(s) \equiv Q_a(s_0) + Q_a'(s')
\end{equation}
Following now a procedure completely analogous to the PQ case, it is
straightforward to verify that the equations for PV buses~\eqref{eq:PW_PV} are
also invariant under the transformation given by~\eqref{eq:renorm} and
\eqref{eq:renormQ}. In this case the new parameters are:
\begin{equation*}
  \begin{split}
    Y_a^{\prime\text{sh}} &\equiv (1-s_0) |V_a(s_0)|^2 Y_a^\text{sh}  \\
    \Gamma'_a &\equiv \Gamma_a + s_0 P_a - j Q_a(s_0)\\
    P_a' &\equiv (1-s_0) P_a
  \end{split}
\end{equation*}

\subsection{Full AC system: complementarity constraints}
Let us now analyze how the complementarity constraint equations
\eqref{eq:compl_embed} transform under the change given by~\eqref{eq:renorm},
\eqref{eq:renormQ}. Consider the most general case, where limits and
setpoints are embedded and voltage regulation may be remote, as described in
Section~\ref{sec:HELMscheme}:
\begin{equation}
  \label{eq:compl_embed2}
  V_b(s) V_b^*(s^*) - W_b^\text{sp}(s) =
  - \frac{\mu_a(1-s)}{Q_a^\text{max}(s) - Q_a(s)}
  + \frac{\mu_a(1-s)}{Q_a(s) - Q_a^\text{min}(s)}
\end{equation}
It is readily verified that this equation is also invariant under the change, if
one defines the new transformed parameters as follows:
\begin{equation*}
  \begin{split}
    \mu_a' &\equiv \frac{(1-s_0)\,\mu_a}{\left|V_b(s_0)\right|^2} \\
    W_b^{\prime\text{sp}}(s') &\equiv
    \frac{W_b^\text{sp}}{\left|V_b(s_0)\right|^2}
    + \frac{(1-s_0)\,\delta W_b^\text{sp}}{\left|V_b(s_0)\right|^2} (1-s') \\
    Q_a^{\prime\text{min}} (s') &\equiv
    Q_a^\text{min} - Q_a(s_0) + \delta Q_a^\text{min} (1-s_0) (1-s') \\
    Q_a^{\prime\text{max}}(s') &\equiv
    Q_a^\text{max} - Q_a(s_0) + \delta Q_a^\text{max} (1-s_0) (1-s')
  \end{split}
\end{equation*}

For the transformed problem, the consistency requirements at $s'=0$ are
automatically satisfied, as the corresponding equation is equivalent to the
original one when $s=s_0$, and $s_0$ is interior. Namely, one has
$Q_a^{\prime\text{min}}(s'=0)<0$, $Q_a^{\prime\text{max}}(s'=0)>0$, and:
\begin{equation*}
  \sgn\left(1-W_b^{\prime\text{sp}}(s'=0) \right) =
    \sgn\left( Q_a^{\prime\text{min}}(s'=0) + Q_a^{\prime\text{max}}(s'=0) \right)
\end{equation*}
This last condition may be satisfied even with a value of zero, i.e.\
$W_b^{\prime\text{sp}}(s'=0)=1$; this poses no problem since the transformed
coefficients $\mu_a'$ are already known. Therefore after the first P-W stage
there is no need to introduce any additional guards in the embedding of neither
the setpoint nor the limits.

This completes then the P-W procedure for the full AC system including Q-limits.
At each stage, one obtains (using standard HELM) a \emph{partial solution} at a
value $0<s_0<1$ such that the required precision is satisfied (which is always
possible, by Stalh). Then one uses this partial solution in $s$-space in order
to define a transformed problem in $s'$-space, using~\eqref{eq:renorm},
\eqref{eq:renormQ}. The transformed problem turns out to be formally the same
as the original one, so that the whole procedure can be repeated recursively,
until one is able to reach the required precision at $s^\text{(k)}=1$ at some
P-W stage $k$. The solution to the original problem is then obtained by undoing
all the transformations defined by using~\eqref{eq:renorm}, \eqref{eq:renormQ}.
In effect, this is a path-analytic continuation in the spirit of Weierstrass,
but the re-expansion of the power series at intermediate points is obtained
using the exact function values (within machine precision, using Padé
approximants) instead of the initial power series.

\subsection{Full AC system: HELM scheme}
It only remains to see how the HELM scheme is slightly modified by the presence
of the new terms $\Gamma$ introduced by the P-W procedure. This only affects the
main power flow equations; the complementarity constraint equations remain exactly
the same as in the first stage, so that the $N$-th order representation and the
resulting HELM scheme are exactly the same as those shown in
Section~\ref{sec:HELMscheme}.

From the equation for PQ buses, \eqref{eq:PW_PQ}, one obtains the following
relationship for the power series coefficients at order $N$ (the so-called
$N$-th order representation):
\begin{multline*}
  \sum_{j=0}^n Y_{ij} V_j[N] + \Gamma_i \left( V_i[N] - V_i^{-1*}[N] \right) = \\
  S_i^* V_i^{-1*}[N-1] - Y_i^\text{sh} V_i[N-1]
\end{multline*}
Recall that the coefficients of $1/V(s)$, $V_i^{-1}[N]$, can be obtained in
terms of those of $V(s)$ by~\eqref{eq:invSeries}. Initial values are
$V_i[0] = V_i^{-1}[0] = 1$. Therefore:
\begin{equation*}
  \sum_{j=0}^n Y_{ij} V_j[N] + 2\Gamma_i \Re V_i[N] = \mathcal{R}_i[N-1]  
\end{equation*}
where all terms on the right hand side depend only on coefficients of order
$N-1$ or less:
\begin{multline*}
  \mathcal{R}_i[N-1] \equiv
  S_i^* V_i^{-1*}[N-1] - Y_i^\text{sh} V_i[N-1] \\ 
  - \Gamma_i \sum_{m=1}^{N-1} V_i^*[m] V_i^{-1*}[N-m]  
\end{multline*}

Analogously, from the equation for PV buses, \eqref{eq:PW_PV}, one obtains:
\begin{multline*}
  \sum_{j=0}^n Y_{aj} V_j[N] + \Gamma_a \left( V_a[N] - V_a^{-1*}[N] \right) = \\
  P_a V_a^{-1*}[N-1] - Y_a^\text{sh} V_a[N-1] - j \sum_{m=0}^N Q_a[m] V_a^{-1*}[N-m]
\end{multline*}
Making use of initial values $V_a^{-1}[0]=1,\, Q_a[0]=0$, and moving all terms
of order $N$ to the left hand side, one obtains:
\begin{equation*}
  \sum_{j=0}^n Y_{aj} V_j[N] + 2\Gamma_a \Re V_a[N] + jQ_a[N] = \mathcal{R}_a[N-1]  
\end{equation*}
where all terms on the right hand side depend only on coefficients of order
$N-1$ or less:
\begin{multline*}
  \mathcal{R}_a[N-1] \equiv
  P_i V_a^{-1*}[N-1] - Y_a^\text{sh} V_a[N-1] \\ 
  - \Gamma_i \sum_{m=1}^{N-1} V_a^*[m] V_a^{-1*}[N-m]
  - j \sum_{m=1}^{N-1} Q_a[m] V_a^{-1*}[N-m]
\end{multline*}
and this concludes the description of the method. Recall that variables $Q_a[N]$
can be eliminated from the constraint equation~\eqref{eq:Nth_constraint2}, thus
leading to a linear system of dimension $2(\Npq+\Npv)$ in the variables
$\Re V[N]$, $\Im V[N]$.

%%%\subsection{Constant-current injections}
%%%\label{subsec:ccInjections}
%%%(TODO)

\section{Numerical results}
\label{sec:NumResults}

\subsection{Improved accuracy and numerical stability near collapse}

Before numerical results are presented on power flow cases with Q-limits, let us
analize the effects of the P-W technique on the base HELM method, when no Mvar
limits are enforced. In order to do so, let us consider the kind of situations
under which the method may suffer from numerical precision problems. There are
two possible sources: very close proximity to the feasibility boundary (voltage
collapse), and ill-conditioning of the transmission admittance matrix. The last
case may occur when some anomalous branch admittances are much larger or much
smaller than the rest, ending up in a HELM linear system whose matrix has a bad
condition number. This will produce errors in the power series coefficients, and
eventually slower convergence. Just as it is done in iterative methods, this is
simply prevented by requiring the admittances to be within certain minimum and
maximum thresholds.

The other case has a more fundamental cause. A case that is feasible but very
near the critical point is a case for which the embedded HELM problem has a
singularity at $s_\text{crit}=1+\varepsilon$, where $0<\varepsilon<<1$. The
point $s_\text{crit}$ marks the transition, as the Padé approximants will start
to oscillate for $s>s_\text{crit}$.  The problem is that, even though Stalh's
theorem still guarantees that the Padé approximant sequences converge at $s=1$
for this case, the convergence rate slows down so much that too many orders of
the power series are needed, and the approximants can no longer be computed with
precision due to the inherent limits of floating point arithmetic. In simpler
terms: as a case approaches the feasibility boundary, the method hits a ceiling
of maximum attainable numerical precision.

In order to quantify this effect, and show how the new P-W technique helps in
mitigating it, some numerical results are presented below. The important point
here is not so much the ability of P-W to obtain ``nose points'' with much
greater precision (after all, it can be shown that the attainable precision of
the base HELM method is sufficient for most practical purposes).  Rather, the
aim here is to provide insight as to why P-W is needed when incorporating
limits, since in that case the complementarity constraints always produce a
singularity right at $s_\text{crit}=1$.

\begin{figure}[!t]
  \centering
  \includegraphics[width=0.98\columnwidth]{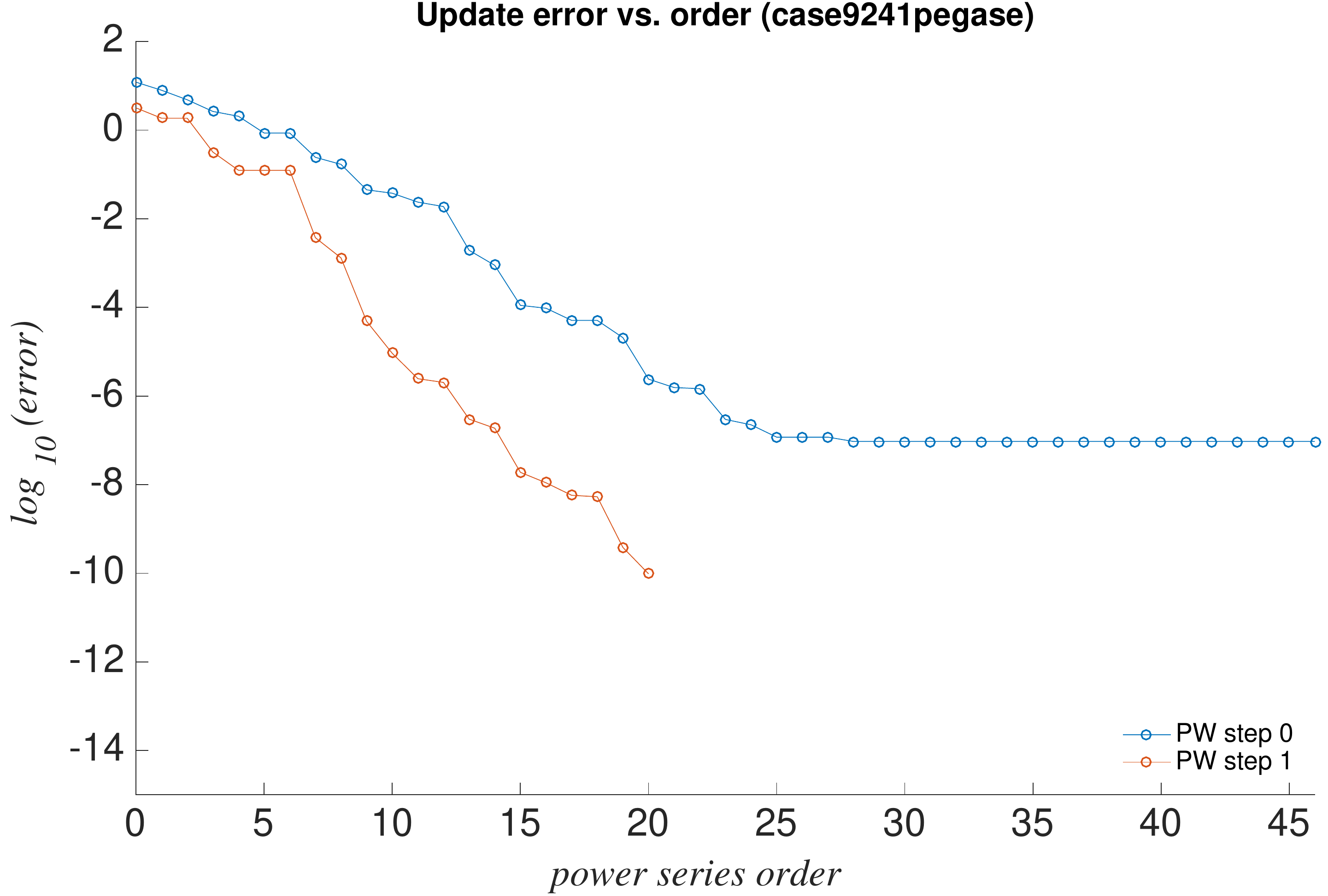}
  \caption{Convergence rates for \texttt{case9241} (unmodified, no Mvar limits
    enforced).  Maximum attainable precision improves from about
    $\pm 10^\text{-7}$ to $\pm 10^\text{-11}$ pu by performing just one P-W
    step.}
  \label{fig:max_attainable_precision}
\end{figure}

Figure~\ref{fig:max_attainable_precision} shows the concept of maximum
attainable precision in an actual example, \texttt{case9241pegase} available in
MATPOWER~\cite{MATPOWER}.  All calculations are performed using standard IEEE
double precision throughout, i.e. about 15 digits of precision.  The figure
plots the overall convergence rate by looking at the maximum update error in
voltages across all buses versus $N$, the number of power series terms. This
update error is calculated as the difference between two successive Padé
approximants down the staircase sequence of the Padé table, e.g. $[L/M]$ and
$[L+1/M]$ (in the notation of~\cite{BakerGraves96}), where one normally uses
$L=M=N/2$. The results show that in this case the standard HELM algorithm is not
able to obtain the voltages with more precision than about $\pm 10^{-7}$ pu, no
matter how many series terms are obtained. This happens because at around 25
terms the evaluation of Padé approximants starts suffering from the limitations
of floating point arithmetic (catastrofic cancellations and
round-off). Performing just one step of the P-W procedure, the new maximum
achievable precision in the transformed problem rises to $\pm 10^{-11}$ pu. A
second P-W step is sufficient to achieve about $\pm 10^{-15}$, thus exausting
the machine precision.

\begin{figure}[!t]
  \centering
  \includegraphics[width=0.98\columnwidth]{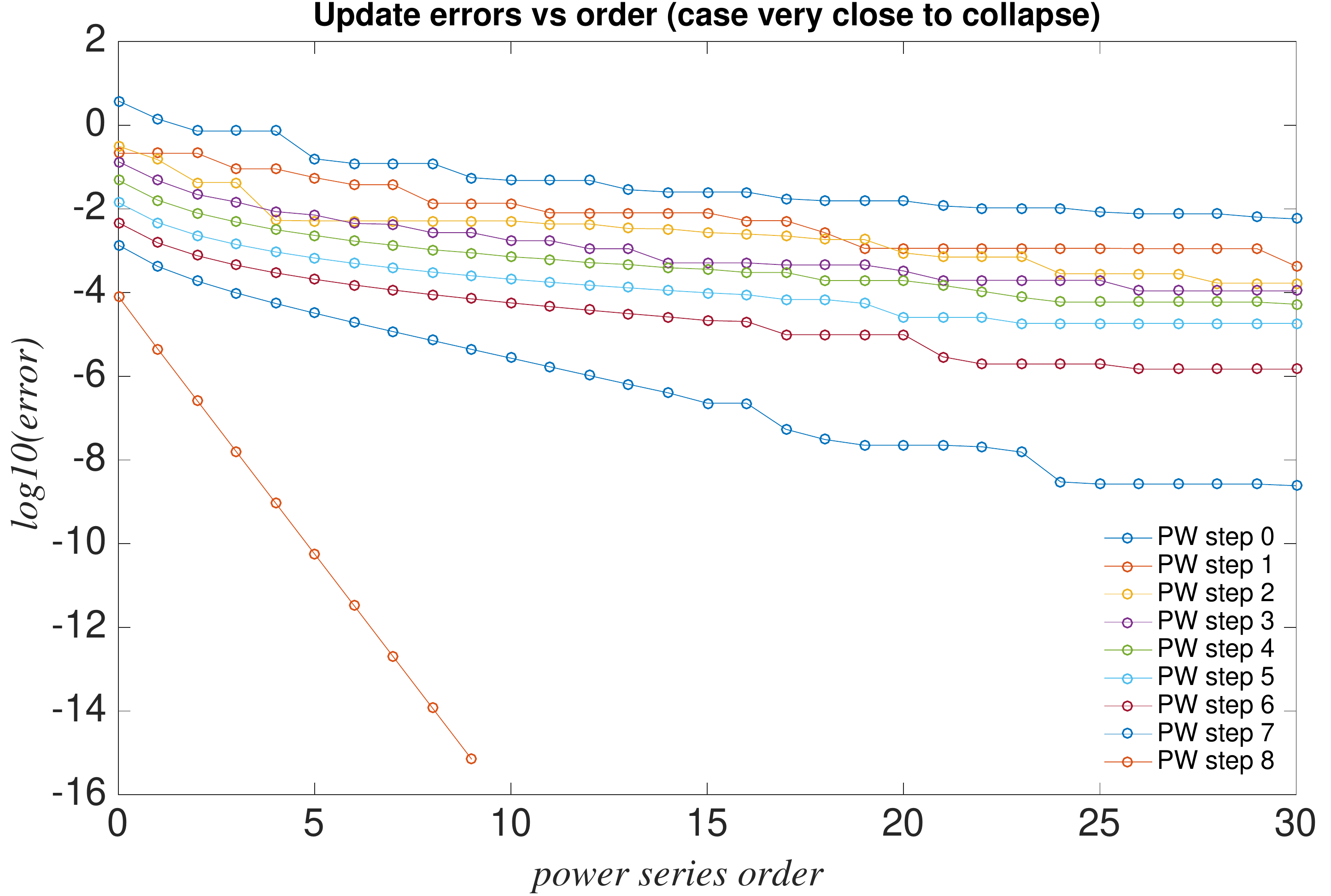}
  \caption{Convergence rates for \texttt{case9} (no Mvar limits enforced), where
    load and generation have been scaled up as close as possible to the point of
    collapse. Maximum attainable precision improves from about
    $\pm 10^\text{-2}$ to $\pm 10^\text{-15}$ pu by performing 8 P-W steps.}
  \label{fig:case9collapse}
\end{figure}

Figure~\ref{fig:case9collapse} demonstrates the effect of successive
applications of the P-W transformation in a more extreme case. The calculation
was performed using \texttt{case9} from MATPOWER, in which a uniform scaling was
applied to all loads and generators in order to bring the system as close as
possible to collapse. The location of the critical point was narrowed down to a
scaling factor of $\lambda_\text{crit}=2.48539267\pm 1.0e-8)$, which was
confirmed using both HELM and MATPOWER's CPF method. Since the total load at the
critical point is around 315 MW, this error margin in the determination of the
exact value of $\lambda_\text{crit}$ corresponds to about 3 Watts.  Therefore,
since Mvar limits are not enforced here, the limiting singularity
$s_\text{crit}$ is very close to, but slightly above 1. At stage zero (base HELM
method), it can be observed that the convergence rate is quite slow and the
maximum achievable precision seems to be around $\pm 10^{-2}$ pu. After 8 P-W
steps, the convergence rate is so fast that just 9 terms are enough to achieve
maximum machine precision.

This behavior can be qualitatively understood in terms of the geometry of the
P-W transformation in $s$-space. Let us recall the parameter transformation,
rewriting~\eqref{eq:renorm} in this form:
\begin{equation*}
  s' = \frac{s - s_0}{1-s_0}
\end{equation*}
It is easy to see that this conformal mapping is a translation and a dilation of
$s$-space, which has the effect of advancing and ``zooming in'' towards the
point $s=1$. Using conformal maps in order to push unwanted singularities away
and thus improve the convergence properties of Padé approximants is a well-known
technique that has been applied to general cases~\cite{BrunoReitich94}. Here it
will be shown how this particular transformation works in the HELM problem.

\begin{figure}[!t]
  \centering
  \includegraphics[width=0.95\columnwidth]{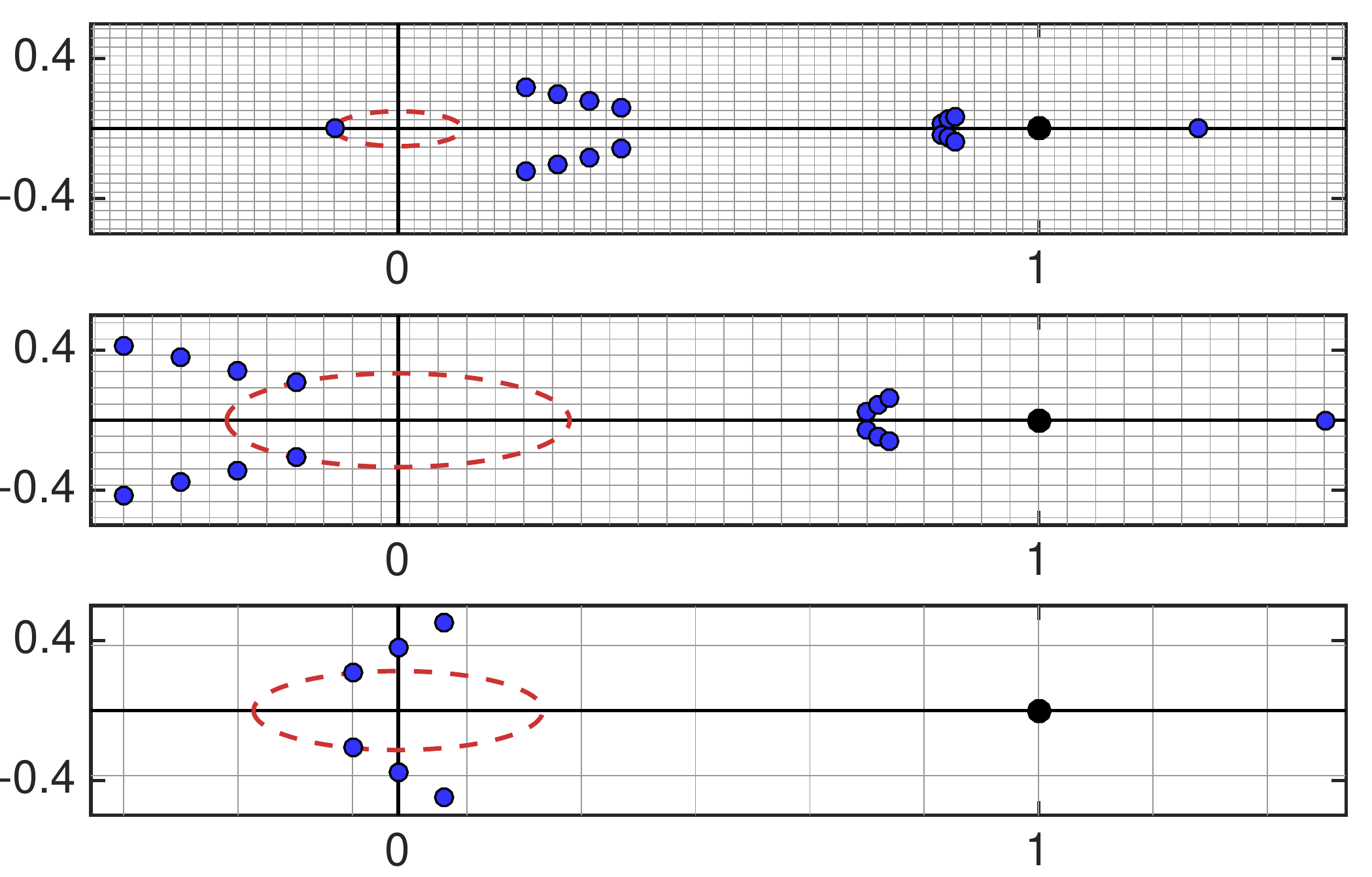}
  \caption{How the P-W transformation acts in $s$-space. The graphs show a set
    of singularities in: the original $s$-space (top); the $s'$-transformed
    space, for $s_0=0.44$ (center); the $s''$-transformed space, for $s'_0=0.75$
    (bottom). A grid is shown to aid the eye in appreciating the dilating
    effects of successive P-W transformations. After the second P-W step, the
    Padé approximants are much better conditioned to converge for $s''=1$.}
  \label{fig:s-space_transform}
\end{figure}

Figure~\ref{fig:s-space_transform} shows graphcally the effect that the
successive P-W transformations have on the singularities of the problem: all
non-relevant singularities get pushed away, eventually leaving the only one that
matters: the first one to be encountered on the positive real
axis~\cite{Trias15}. When this critical singularity is $s_\text{crit}>1$, the
successive P-W transformations will keep pushing it away until it is so far that
the power series is sufficienly well-conditioned to allow its Padé approximants
to converge quickly (one could even obtain a series whose radius of convergence
included 1, although this is not strictly necessary). If on the other hand
$s_\text{crit}<1$, the P-W procedure would ``pull in'' the singularity towards 0
and quickly get stuck (the successive values $s_0$ would approach zero), thus
signaling infeasibility. As a by-product, this would also provide a precise
value for $s_\text{crit}$. Finally, if $s_\text{crit}=1$ then all other
singularities are pushed away but $s_\text{crit}$ will always remain at a
distance of 1. The Padé approximants of every HELM subproblem will still
\emph{not} converge exactly at 1, but all that is needed is that they converge
\emph{near} 1.  Then, thanks to the zooming property, one can get extremely
close to 1 \emph{in the original} $s$-space. For instance, assume a case where
one always obtains sufficient convergence at $s_0=0.9$ in every P-W stage. At
the second stage, the starting point $s''=0$ represents $s=0.99$. After 10 P-W
stages, the partial solution will be just $10^{-10}$ away from 1 in the original
$s$-space. This qualitatively explains the good numerical results that are
obtained in the next section.

\subsection{Examples with Q-limits}

\begin{figure}[!t]
  \centering
  \includegraphics[width=0.98\columnwidth]{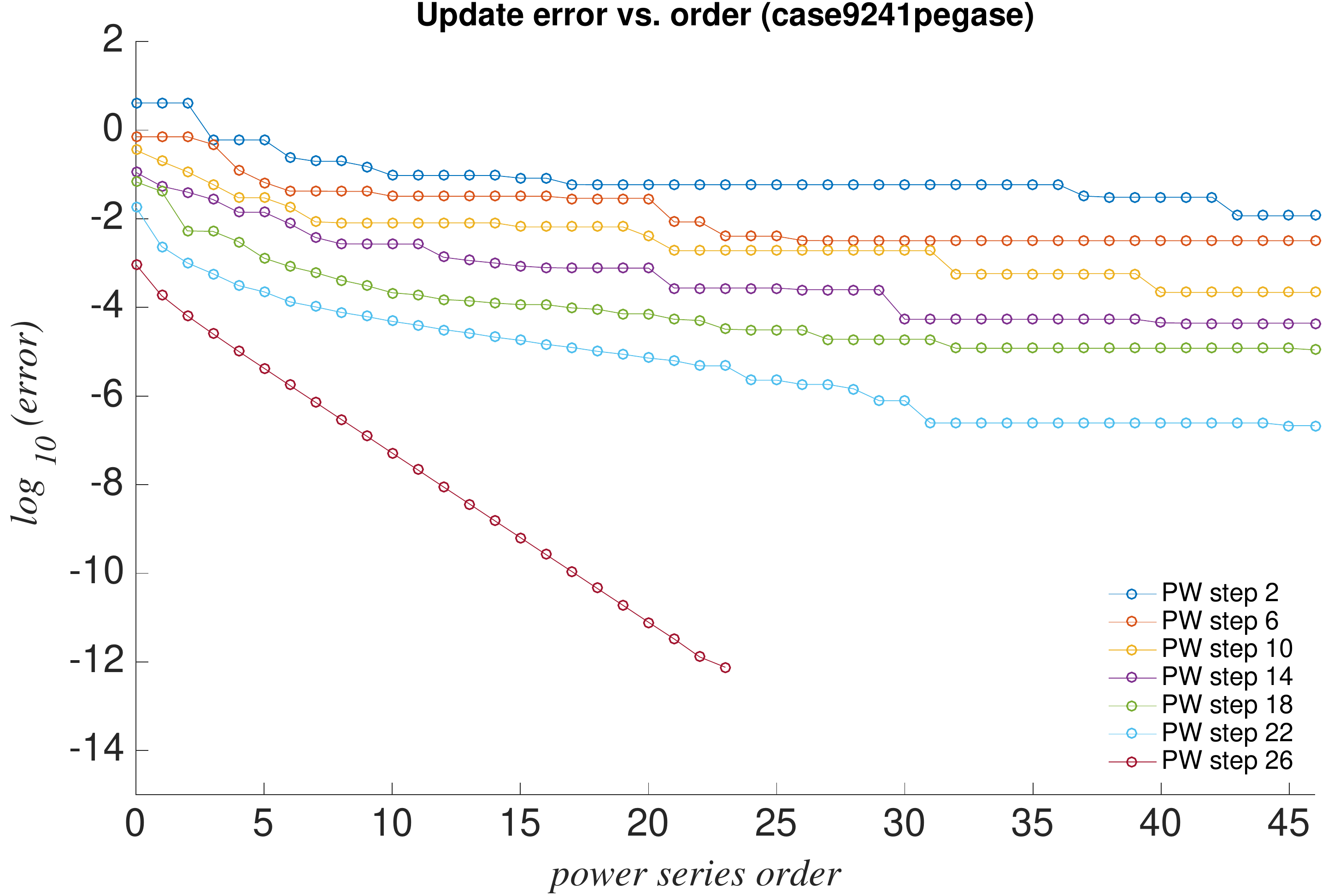}
  \caption{Convergence rates for \texttt{case9241} (unmodified, with Mvar limits
    enforced).  A precision of about $\pm 10^\text{-12}$ pu is achieved ater 26
    P-W steps.}
  \label{fig:case9241pegase_withQlims}
\end{figure}

Figure~\ref{fig:case9241pegase_withQlims} shows the convergence rates for
\texttt{case9241pegase} when Mvar limits are enforced using the P-W
method. Contrast this with Figure~\ref{fig:max_attainable_precision}, where
limits are not enforced.  The presence of the complementarity constraints
introduces an unavoidable singularity at $s=1$, whose effects can be clearly
seen in the convergence rates. However, the application of the P-W technique
achieves the solution with good precision.

Table~\ref{tab:PWresults} shows the results of applying the HELM-PW method to
the top 25 largest cases available in MATPOWER. The numerical tolerances
required for all cases were $10^{-11}$ pu maximum update error in voltages (at
each P-W stage), and $10^{-8}$ pu maximum mismatch error in the original power
flow equations (nodal currents). The maximum number of power series terms used
at each P-W stage was $N_\text{max}=32$. Note that in all these examples
regulation is always local to the bus, and that $Q^\text{min},Q^\text{max}$
represent the aggregate limits of all generators connected to the bus, in case
there is more than one.

The first column shows the number of P-W stages needed in order to meet the
required tolerance; this provides a rough idea of the numerical difficulty
involved in arriving to the solution.  The next three columns show the number of
PV buses that end up saturated at either $Q^\text{min}$ or $Q^\text{max}$,
comparing the results of HELM vs.\ the two limit-enforcing strategies available
in MATPOWER's Newton-Raphson method (strategy 1 consists in type-switching all
violated buses at once, on every NR solution; while strategy 2 consists in
type-switching one bus at a time). One can observe how there are many cases
where the solutions obtained with MATPOWER are different, and a few cases where
NR does not converge. Additionally, HELM's solutions seem to consistently yield
a lower total number of saturated buses.
 
The table also shows the percentage of saturated buses over the total number of
PV buses. Interestingly, there seems to be no clear correlation between this
number and the number of P-W steps required, which probably indicates that the
numerical difficulty of the case may be influenced by other factors, such as
proximity to voltage collapse. This issue has been left for further research.

\setlength\tabcolsep{5pt} % To avoid overfull box from this table (default value: 6pt)
\begin{table}[t]
  \centering
  \begin{tabular}{lccccr}
    \toprule
    & P-W & \multicolumn{3}{c}{sat.\ $Q^\text{min}/Q^\text{max}$} \\
    \cmidrule(lr){3-5}
    Case        & steps & HELM    & MP-1             & MP-2            & \% PV\\
    \midrule
    128         &  6    & 5/1     &  =               &  =              & 11.3 \\
    145         &  6    & 0/1     &  =               &  =              &  2.0 \\
    illinois200 &  8    & 0/13    &  =               &  =              & 35.1 \\
    300         & 15    & 0/11    &  =               &  =              & 16.2 \\
    300\_PS     & 15    & 0/13    &  =               &  =              & 19.1 \\
    1354pegase  & 14    & 0/25    &  =               &  =              &  9.7 \\
    1888rte     & 23    & 13/2    & 13/\textbf{3}    &  =              &  5.5 \\
    1951rte     & 22    & 4/9     & \textbf{8/7}     &  =              &  3.7 \\
    2383wp      & 16    & 149/222 & \textbf{NC}      & \textbf{NC}     & 75.8 \\
    2736sp      & 19    & 67/99   & \textbf{73/102}  &  =              & 53.6 \\
    2737sop     & 23    & 63/87   & \textbf{NC}      &  =              & 56.8 \\  
    2746wop     & 23    & 206/155 & \textbf{NC}      & \textbf{NC}     & 76.8 \\
    2746wp      & 18    & 98/235  & \textbf{107/239} & \textbf{99}/235 & 67.2 \\
    2848rte     & 15    & 27/32   & \textbf{30/34}   &  =              & 15.7 \\
    2868rte     & 13    & 12/31   & \textbf{20/33}   & \textbf{13}/31  & 10.2 \\
    2869pegase  & 19    & 0/72    &  =               &  =              & 14.1 \\
    3012wp      & 19    & 112/194 & \textbf{113}/194 &  =              & 66.0 \\
    3120sp      & 15    & 124/143 & \textbf{NC}      & \textbf{NC}     & 67.6 \\
    3375wp      & 18    & 117/182 & \textbf{118/184} &  =              & 48.3 \\  
    6468rte     & 53    & 7/40    & \textbf{10/38}   &  =              & 16.2 \\
    6470rte     & 38    & 34/38   & \textbf{36/37}   &  =              & 15.5 \\ 
    6495rte     & 36    & 16/56   & \textbf{22/53}   &  =              & 14.6 \\
    6515rte     & 60    & 11/68   & \textbf{14/66}   &  \textbf{12}/68 & 16.0 \\
    9241pegase  & 20    & 6/190   & \textbf{7}/190   &  =              & 13.6 \\
    13659pegase & 10    & 0/1     &  =               &  =              & 0.02 \\
    \bottomrule
  \end{tabular}
  \caption{Numerical results of the P-W method applied to the top 25 largest cases
    available in MATPOWER. The MP-1 and MP-2 columns report the MATPOWER results
    when using NR with the two alternative limit-enforcing strategies
    (\texttt{enforce\_q\_lims} 1 and 2, respectively). Results equal to
    HELM's are marked with (=), differences are
    highlighed in bold, and non-convergence shown as \textbf{NC}.}
  \label{tab:PWresults}
\end{table}

\subsection{Some preliminary results on performance}
A performance evaluation of the original HELM method, which did not have Q-limit
enforcement, was presented in~\cite{TriasMarin16}, using a MATLAB-based
implementation and comparing results to several methods available in
MATPOWER. Since our current implementation of the P-W method has not yet
undergone a proper optimization, its performance will not be reviewed
here. Instead, some scaling arguments will be given in order to roughly
characterize the expected performance.

As seen in table~\ref{tab:PWresults}, it is observed that the method requires a
number of P-W steps that varies between 6 and 60, for this sample of cases and a
choice of $N_\text{max}=32$. Since each P-W step involves roughly one full HELM
computation, this provides an approximate idea of the kind of performance one
may expect. There is a certain trade-off between $N_\text{max}$ and the
resulting number of P-W steps needed to achieve the required tolerance. If one
uses a large value for $N_\text{max}$ (e.g., more than 60 power series terms),
the advancing values $s_0$ at each P-W step will be larger, and therefore the
total amount of P-W steps needed will be lower. However, the HELM problem being
solved at each of those steps will be more costly, because there will be more
series terms to solve.  This would involve solving more linear systems, with
their corresponding right-hand sides (which involve convolutions); and also
computing Padé approximants at larger orders.  The authors have not yet tried to
find rigorously the optimum point for this trade-off, but initial numerical
results indicate that $N_\text{max}\approx 30$ gives good results.

\section{Conclusions and further research}

A new HELM-based method has been presented that deals with control limits in a
rigorous way, based on a novel Lagrangian formulation of the power flow
equations.  The method has been developed in full detail for the case of Mvar
limits of voltage-controlling generators in AC networks, which is the most
important problem of this type.  Like HELM, the method is direct, constructive,
and deterministic.

The Lagrangian approach allows framing the problem in terms of constrained
minimization, leading to a HELM scheme for calculating that minimum. Two
additional key ingredients are also necessary: on the one hand, a proper
embedding for setpoints and Q-limit values, in order to ensure consistency of
the HELM reference state at $s=0$. On the other, the application of the P-W
technique, which enables the possibility of calculating the analytic
continuation of the reference state to $s=1$ with good numerical precision.  P-W
exploits a remarkable property of the power flow equations, whereby certain
changes of variables and the embedding parameter leave the equations formally
invariant. This allows one to calculate the analytic continuation of the power
series in several steps, leveraging the power flow equations at each step to
re-expand new auxiliary series. The net result is an analytic continuation along
a path, with far greater numerical stability.

In contrast to traditional iterative methods, this one directly avoids the
convergence problems derived from mutual interactions among controls, which can
produce oscillatory behavior, particularly when several nearby controls are
close to their saturation limits. The method presented here is closer in spirit
to some OPF-like approaches to power flow, although the differences are
fundamental, since HELM is based on algebraic curves and complex analysis.

The authors have also applied this method to other types of regulating devices
with limits, in the context of DC microgrids~\cite{TriasMarin16}. Examples are
the sequential shunt unit (SSU) that regulates the output voltage of solar
panels aboard spacecraft, and the voltage regulation of DC-DC converters. In
those cases the regulation can also be expressed in terms of algebraic
constraints involving new regulating injections, and their limits can be
accommodated under the same P-W technique even though they are sometimes
dependent on voltage.

Future areas to explore include the extensions to other controls, such as
automatic tap changers (ULTC transformers), automatic phase-shifting
transformers, or inter-area flow constraints. Shared control of a given point
originated from different buses is also a challenging goal.  On the other hand,
the resemblance between the method presented here and certain OPF methods
certainly calls for a more detailed study on the possible applications to
optimization problems.

% Criticism: Other OPF-like methods for PF still relying on predictor-corrector (invariably
% depend on NR; sometimes NR falls onto wrong solution branch, Kataoka
% warnings).

% Our helm method can be used as a way to tell feasibility, more efficient than
% optimization-based methods such as ~\cite{Molzahn13}.

\appendix

\section{Method summary}
\label{app:methsummary}
The HELM P-W method is summarized here for the general case corresponding to a
set of $\Npq$ buses of type PQ (labeled with indexes $i$ when the distinction is
needed) and $\Npv$ buses (labeled with indexes $a$) having a reactive power
injection that controls the voltage at a corresponding set of buses (labeled
with indexes $b$). A bus where $a=b$ is typically referred to as being PV-type.
It is assumed that there are no concurrent controls, i.e.\ each controlled bus
$b$ has only one corresponding controlling injection $a$. The injections are
assumed to have both upper and lower limits: $Q_a^\text{min}, Q_a^\text{max}$.
In case of one-sided limits, the changes in the formulation are all
straightforward, except for the special care that must be taken in embedding the
limits and the setpoint in the complementarity equation (see
examples~\eqref{eq:JLMembeddingconstants1limmin} and
~\eqref{eq:JLMembeddingconstants1limmax} in Section~\ref{sec:HELMscheme}).

\subsection*{Embedding and reference state}
The embedded power flow equations (see Section~\ref{sec:theproblem} for notation
conventions): \begin{equation*}
  \begin{split}
   \sum_{j=0}^n Y_{ij} V_j(s) + sY_i^\text{sh} V_i(s) &=
    \frac{sS_i^*}{V_i^*(s^*)} + \Gamma_i \left(\frac{1}{V_i^*(s^*)} - V_i(s)\right)\\
    \sum_{j=0}^n Y_{aj} V_j(s) + sY_a^\text{sh} V_a(s) &=
    \frac{sP_a - jQ_a(s)}{V_a^*(s^*)} +
    \Gamma_a \left(\frac{1}{V_a^*(s^*)} - V_a(s)\right)
  \end{split}
\end{equation*}
where all $\Gamma$ parameters are zero at the first P-W stage. The swing is
embedded as $V_0(s)=1+s(V_0-1)$, so as to allow a reference state having
$V_j[0]=1$ for all $j$. Reactive injection variables are $Q_a[0]=0$ for all $a$.

Reactive limits are contemplated by means of the embedded complementarity
equation: \begin{equation*}
  V_b(s) V_b^*(s^*) - W_b^\text{sp}(s) =
  - \frac{\mu_a(1-s)}{Q_a^\text{max}(s) - Q_a(s)}
  + \frac{\mu_a(1-s)}{Q_a(s) - Q_a^\text{min}(s)}
\end{equation*}
where $W_b^\text{sp}\equiv V_b^\text{sp} V_b^{\text{sp}*}$. Limits and setpoints
need to be adequately embedded in order to guarantee certain consistency
requirements as described in Section~\ref{sec:HELMscheme}. The following choice
is recommended:
\begin{equation*}
  \begin{split}
    \mu_a &= 1 \\
    Q_a^\text{min}(s) &= Q_a^\text{min} - \frac{Q_a^\text{max}+Q_a^\text{min}}{2} (1-s) \\
    Q_a^\text{max}(s) &= Q_a^\text{max} - \frac{Q_a^\text{max}+Q_a^\text{min}}{2} (1-s) \\
    W_b^\text{sp}(s)  &= W_b^\text{sp} +\left( 1 - W_b^\text{sp} \right) (1-s)
  \end{split}
\end{equation*}
so that the coefficients of their corresponding power series are:
 \begin{align*}
    W_b^\text{sp}[0]  &= 1 &
    Q_a^\text{max}[0] &= - Q_a^\text{min}[0] = \frac{Q_a^\text{max}-Q_a^\text{min}}{2} \\
    W_b^\text{sp}[1]  &= \left( W_b^\text{sp} - 1 \right) &
    Q_a^\text{max}[1] &= Q_a^\text{min}[1] = \frac{Q_a^\text{max}+Q_a^\text{min}}{2}
\end{align*}
However, the formulas in the next section will not assume any particular choice
for these embedding constants; only the linearity of the embedding is
assumed.

\subsection*{HELM scheme}
At each stage of the P-W procedure, solve the following HELM equations
sequentially order by order, to obtain the power series of voltages and reactive
injections:
\begin{equation*}
  \begin{split}
    \sum_{j=0}^n Y_{ij} V_j[N] + 2\Gamma_i \Re V_i[N] = \mathcal{R}_i[N-1]  \\
    \sum_{j=0}^n Y_{aj} V_j[N] + 2\Gamma_a \Re V_a[N] + jQ_a[N] = \mathcal{R}_a[N-1]
  \end{split}
\end{equation*}
The terms on the right hand side depend only on coefficients of order $N-1$ or
less, and are calculated as:
\begin{gather*}
  \begin{split}
    \mathcal{R}_i[N-1] \equiv S_i^* V_i^{-1*}[N-1] &- Y_i^\text{sh} V_i[N-1] \\ 
    &- \Gamma_i \sum_{m=1}^{N-1} V_i^*[m] V_i^{-1*}[N-m]
  \end{split}\\
  \begin{split}
    &\mathcal{R}_a[N-1] \equiv P_i V_a^{-1*}[N-1] - Y_a^\text{sh} V_a[N-1] \\ 
    &- \Gamma_i \sum_{m=1}^{N-1} V_a^*[m] V_a^{-1*}[N-m]
    - j \sum_{m=1}^{N-1} Q_a[m] V_a^{-1*}[N-m]
  \end{split}
\end{gather*}
and the coefficients of $1/V(s)$ can be obtained from those of $V(s)$ by:
\begin{equation*}
  V_j^{-1}[N] = - V_j[N] - \sum_{m=1}^{N-1} V_j[m] V_j^{-1}[N-m]  
\end{equation*}
Variables $Q_a[N]$ are to be solved below in terms of $\Re V_b[N]$ from the constraint
equation, and substituted into the above linear system. To do so, first
construct he auxiliary power series $\mathcal{B}_a^\text{(+/-)}$, using:
\begin{equation*}
  \begin{split}
    \mathcal{B}_a^+[0] =& -\frac{\mu_a}{Q_a^\text{max}[0]} \\
    \mathcal{B}_a^+[N] =& \frac{1}{Q_a^\text{max}[0]} \Biggl[
    \sum_{m=0}^{N-1} \mathcal{B}_a^+[m] \, Q_a[N-m] \Biggr. \\
    & \Biggl. - \mathcal{B}_a^+[N-1] \, Q_a^\text{max}[1] + \mu_a\,\delta_{1,N} \Biggr]
  \end{split}
\end{equation*}
and the same expressions for $\mathcal{B}_a^-$, using $Q_a^\text{min}[0]$ and
$Q_a^\text{min}[1]$ instead. Having calculated the coefficients of
$\mathcal{B}_a^\text{(+/-)}$ up to order $N-1$, the coefficients of $Q_a[N]$ are
solved from:
\begin{equation*}
  2\Re V_b[N] - \left( \frac{\mathcal{B}_a^-[0]}{Q_a^\text{min}[0]}
              + \frac{\mathcal{B}_a^+[0]}{Q_a^\text{max}[0]} \right) Q_a[N] =
  \mathcal{T}_a[N-1]
\end{equation*}
and substituted into the main equations in order to form a linear system of
dimension $2(\Npq+\Npv)$ in the variables $\Re V[N]$, $\Im V[N]$. The expression
for the right hand side in the above expression is:
\begin{multline*}
  \mathcal{T}_a[N-1] \equiv W_b^\text{sp}[1]\,\delta_{1,N} \,
  - \sum_{m=1}^{N-1}V_b[m]\,V_b^*[N-m] \\
  + \frac{1}{Q_a^\text{min}[0]} \Biggl[ \sum_{m=1}^{N-1} \mathcal{B}_a^-[m] Q_a[N-m]
  - \mathcal{B}_a^-[N-1] Q_a^\text{min}[1] + \mu_a\,\delta_{1,N} \Biggr] \\
  + \frac{1}{Q_a^\text{max}[0]} \Biggl[ \sum_{m=1}^{N-1} \mathcal{B}_a^+[m] Q_a[N-m]
  - \mathcal{B}_a^+[N-1] Q_a^\text{max}[1] + \mu_a\,\delta_{1,N} \Biggr]
\end{multline*}

Note how the matrix of the linear system stays constant across all orders $N$,
so the system needs to be factorized only once (at the beginning of each P-W
stage), and the factors are reused to solve the same linear system with a
different right hand side at each order. Of course, it is strongly recommended
to use a modern sparse LU solver~\cite{Davis06}, for efficiency. For best
results, we recommend using either CXSparse~\cite{Davis06} or KLU~\cite{KLU},
together with the Approximate Minimum Degree reordering
algorithm~\cite{Amestoy96} for reducing factor fill-in.

Set a maximum number of power series terms $N^\text{max}$ somewhere between 20
to 40. At each order (or maybe after the first ten terms, to save some work in
most cases), evaluate the corresponding Padé appproximants at $s=1$. If
convergence within the desired tolerance is obtained, the solution has been
found. If $N^\text{max}$ is reached without convergence, find an intermediate
value $0<s_0<1$ such that convergence of the partial solution $V_j(s_0)$ is
obtained (this is guaranteed by Stahl). The maximum achievable $s_0$ can be
found by a simple bisection procedure or any other similar method. Also, one can
select a value of $s_0$ slightly smaller than the maximum possible, thus trading
smaller P-W steps for slightly better overall numerical stability.

\subsection*{Padé-Weierstrass updates}
The partial solution $V_j(s_0), Q_a(s_0)$ is now used to construct the
transformed HELM problem of the next P-W stage. The embedding parameter and the
variables transform as:
\begin{equation*}
  \begin{split}
    s &\equiv s_0 + (1-s_0)s' \\
    V_j(s) &\equiv V_j(s_0) V'_j(s') \\
    Q_a(s) &\equiv Q_a(s_0) + Q_a'(s')
  \end{split}
\end{equation*}
This change leaves all the embedded equations, both the power flow and the
complementarity constraints, \emph{invariant}. One should keep track of all
these transformations in order to undo them once the point $s^{(k)}=1$ can be
reached at the $k$-th P-W stage, when numerical tolerances are met.

For the power flow system, the new parameters are as follows. The new
transmission admittance matrix is given by:
\begin{equation*}   
  Y'_{ij} \equiv
  \begin{cases}
    \hat{Y}_{ij}                           & \text{if } i \ne j \\
    \hat{Y}_{ii} - \sum_{l=0}^n \hat{Y}_{il}  & \text{if } i=j 
  \end{cases} \quad \text{where: } \hat{Y}_{ij} \equiv V^*_i(s_0) Y_{ij} V_j(s_0)
\end{equation*}
and where both indices $i$ and $j$ run over all buses, including the swing. The
new shunt admittances are, also for all buses:
\begin{equation*}
  Y_j^{\prime\text{sh}} \equiv (1-s_0) |V_j(s_0)|^2 Y_j^\text{sh}
\end{equation*}
Now the power and gamma terms are as follows, depending on the type of bus:
\begin{align*}
  \Gamma'_i &\equiv \Gamma_i + s_0 S_i^* & \Gamma'_a &\equiv \Gamma_a + s_0 P_a - j Q_a(s_0)\\
    S_i' &\equiv (1-s_0) S_i & P_a' &\equiv (1-s_0) P_a
\end{align*}
Finally, the new parameters in the complementarity constraint equation are as
follows:
\begin{equation*}
  \begin{split}
    \mu_a' &\equiv \frac{(1-s_0)\,\mu_a}{\left|V_b(s_0)\right|^2} \\
    W_b^{\prime\text{sp}}[0] &\equiv
    \frac{W_b^\text{sp}[0]+s_0 W_b^\text{sp}[1]}{\left|V_b(s_0)\right|^2} \\
    W_b^{\prime\text{sp}}[1] &\equiv
    \frac{(1-s_0)W_b^\text{sp}[1]}{\left|V_b(s_0)\right|^2} \\
    Q_a^{\prime\text{lim}}[0] &\equiv Q_a^\text{lim}[0] + s_0Q_a^\text{lim}[1] - Q_a(s_0) \\
    Q_a^{\prime\text{lim}}[1] &\equiv (1-s_0)Q_a^\text{lim}[1]
  \end{split}
\end{equation*}

%% \section{Unstable PV buses}
%% \label{app:unstablePV}
%% \ttfamily
%% \noindent
%% TODO, MAYBE: Briefly describe the phenomenon of unstable PV buses. Q-V curve
%% and the point
%% of tangency with voltage collapse.  Why HELM provides them as solutions even though
%% they are strictly speaking unstable. Why they should be eliminated (i.e.\
%% corrected) in order for the Padé-Weierstrass method (for limits) to work
%% correctly.
%% \rmfamily

% Any acknowledgements go here, right before the References
%
% From the Author Guide: "If no funding has been provided for the research,
% please include the following sentence: This research did not receive any
% specific grant from funding agencies in the public, commercial, or
% not-for-profit sectors."

\section*{References}
\bibliography{./bib/IEEEabrv,./bib/HELM}

\end{document}